\documentclass[]{pasj02}
\usepackage[switch,mathlines]{lineno}  
\usepackage{bm}           
\usepackage{graphicx}     
\usepackage{amsmath}      

\jyear{2026}
\Received{}
\Accepted{}

\newcommand{\mum}{\thinspace\rm{\mu m}}
\newcommand{\etaO}{\eta_{\rm{O}}}
\newcommand{\etaA}{\eta_{\rm{A}}}

\newcommand{\gmag}{\gamma_{\rm{mag}}}
\newcommand{\gturb}{\gamma_{\rm{turb}}}
\newcommand{\bthm}{\beta_{\rm{thm}}}

\newcommand{\brot}{\beta_{\rm{rot}}}

\newcommand{\msun}{\thinspace M_{\odot}}

\newcommand{\yr}{\thinspace \rm{yr}}  
\newcommand{\au}{\thinspace \rm{au}}  
\newcommand{\gcm}{\thinspace \rm{g} \thinspace \rm{cm}^{-3}}
\newcommand{\gcmcms}{\thinspace \rm{g} \thinspace \rm{cm^2} \thinspace \rm{s}^{-1}}  

\newcommand{\Jcore}{\bm{J}_{\rm{c,net}}}
\newcommand{\magB}{\bm{B}}
\newcommand{\magBhat}{\bm{\hat{B}}}
\newcommand{\magmu}{\mu}

\newcommand{\vel}{\bm{v}}
\newcommand{\rad}{\bm{r}}
\newcommand{\nhat}{\hat{\bm{n}}}

\newcommand{\cms}{\thinspace \rm{cm} \thinspace \rm{s}^{-1}}

\newcommand{\kms}{\thinspace \rm{km} \thinspace \rm{s}^{-1}}
\newcommand{\machs}{{\cal{M}}_{\rm{s}}}
\newcommand{\machA}{{\cal{M}}_{\rm{A}}}
\newcommand{\ssum}{\textstyle\sum\limits}  
\newcommand{\tff}{t_{\rm{ff}}}
\newcommand{\tp}{t_{\rm{p}}}
\newcommand{\tpform}{t_{\rm{0}}}
\newcommand{\tpend}{t_{\rm{p,end}}}
\usepackage{xcolor}
\definecolor{steelblue}{rgb}{0.275, 0.510, 0.706}
\definecolor{seagreen}{rgb}{0.190, 0.525, 0.361}
\definecolor{alcrimson}{rgb}{0.227, 0.38, 0.54}

\begin{document}

\title{
  Competitive roles of turbulence and magnetic field on the formation
  of unipolar outflows in isolated molecular cloud cores
}

\author{
  Daisuke \textsc{Takaishi},\altaffilmark{1}\altemailmark\orcid{0000-0001-6273-805X} \email{takaishi@ibe.kagoshima-u.ac.jp}
  Yusuke \textsc{Tsukamoto},\altaffilmark{1,2}\altemailmark\orcid{0000-0001-6738-676X}
  Miyu \textsc{Kido},\altaffilmark{1,3,4}\altemailmark\orcid{0000-0002-2902-4239}
  Shigehisa \textsc{Takakuwa},\altaffilmark{1,3}\altemailmark\orcid{0000-0003-0845-128X}
  Yoshiaki \textsc{Misugi},\altaffilmark{5}\altemailmark\orcid{0000-0002-4093-6925}
  Yuki \textsc{Kudoh},\altaffilmark{6,7}\altemailmark\orcid{0000-0003-0548-1766}
  and
  Yasushi \textsc{Suto},\altaffilmark{8}\altemailmark\orcid{0000-0002-4858-7598}
}
\altaffiltext{1}{Graduate School of Science and Engineering, Kagoshima University, 1-21-40 Korimoto, Kagoshima, Kagoshima 890-0065, Japan}
\altaffiltext{2}{Faculty of Science and Engineering, Konan University, 8-9-1 Okamoto, Kobe, Hyogo 658-8501, Japan}
\altaffiltext{3}{Academia Sinica Institute of Astronomy and Astrophysics, 11F of Astronomy-Mathematics Building, AS/NTU, No. 1, Sec. 4, Roosevelt Road, Taipei 10617, Taiwan}
\altaffiltext{4}{Star and Planet Formation Laboratory, RIKEN Cluster for Pioneering Research, Wako, Saitama 351-0198, Japan}
\altaffiltext{5}{Faculty of Science and Engineering, Kyushu Sangyo University, 2-3-1 Matsukadai, Fukuoka, Fukuoka 813-8503, Japan}
\altaffiltext{6}{Astronomical Institute, Tohoku University, 6-3 Sendai, Miyagi 980-8578, Japan}
\altaffiltext{7}{Information Technology Center, The University of Tokyo, Kashiwa, Chiba 277-8582, Japan}
\altaffiltext{8}{Research Institute, Kochi University of Technology, Tosa Yamada, Kochi 782-8502, Japan}


\maketitle

\begin{abstract}
  We investigate the impact of turbulence and magnetic field on the
  formation of unipolar outflows in the gravitational collapse of
  isolated low-mass cloud cores using three-dimensional non-ideal
  magnetohydrodynamics simulations.  In the models explored in this
  study, we find that unipolar outflows are formed in cloud cores with
  the super-Alfv\'{e}nic turbulence of $\machA>1$, where $\machA$ is
  the Alfv\'{e}n Mach number of the initial cloud core.  In contrast,
  bipolar outflows are formed in cloud cores with $\machA\lesssim1$.
  Our results suggest that the relative strength of the turbulence to
  the magnetic field in the initial cloud core plays a key role in the
  formation of unipolar outflows.  We also find that no outflow is
  formed in cloud cores with the strong magnetic field of the
  dimensionless mass-to-flux ratio $\magmu=2$.  The outflow
  morphologies can be used as tracers to indirectly estimate the
  relative strengths of the turbulence and magnetic field of the
  parent cloud cores.
\end{abstract}


\section{Introduction}
\label{sec:intro}

Outflows, frequently observed in star-forming regions, play an
important role in the process of star and planet formation.  During
the gravitational collapse of molecular cloud cores, the outflows give
an indication of the birth of the young stellar objects (YSOs)
including the first cores, protostars, and protoplanetary disks (e.g.,
\cite{2002ApJ...575..306T, 2004ApJ...616..266M, 2006ApJ...641..949B,
  2008ApJ...676.1088M}).  The powerful outflows have a significant
dynamical influence on the gas and dust evolution in the
protoplanetary disks via the transport of the mass, linear momentum,
and angular momentum (e.g., \cite{2000ApJ...528L..41T,
  2010ApJ...714L..58T, 2013MNRAS.431.1719M, 2020ApJ...896..158T,
  2020ApJ...900..180M, 2021ApJ...920L..35T, 2023A&A...670A..61M,
  2023MNRAS.521.2661K, 2025ApJ...984..143U}).

Observations of star-forming regions have revealed that the outflows
driven by the YSOs exhibit a morphological diversity.  Of particular
interest is the fact that unipolar outflows, which have only one
directional component that is either redshifted or blueshifted
exclusively, are occasionally observed in star-forming regions (e.g.,
\cite{1993prpl.conf..603F, 2004A&A...426..503W, 2013ApJ...778...72F,
  2018ApJ...863...19A, 2018A&A...618A.120L, 2019ApJ...887..209A,
  2020A&A...634L..12D, 2020A&A...636A..65G, 2020ApJ...903..119L,
  2021ApJ...910...11O, 2023ApJ...944...92S, 2023ApJ...947...25H,
  2024AJ....167...72D}).  \citet{2004A&A...426..503W} conducted a
statistical study of the properties of molecular outflows and found
that 50 out of 391 sources (28 redshifted and 22 blueshifted) exhibit
the unipolar outflows, although 327 sources show the bipolar
structure.  Optical and near-infrared observations have detected
unipolar cavities in the circumstellar envelope delineated by the
scattered light of protostars, which may provide possible evidence for
the unipolar outflows (e.g., \cite{2021ApJ...911..153H}).

The origin of the observed unipolar outflows remains unclear.  One
possible explanation is that they are intrinsically bipolar outflows
that appear unipolar due to the apparent effects of observation, such
as extinction, low sensitivity, and geometrical effects.  For
instance, the unipolar cavities can be affected by the extinction
effect depending on the inclination angle of the outflow
(\cite{2021ApJ...911..153H}).  However, it remains uncertain whether
all the observed unipolar outflows can be explained by the apparent
effects of observation.

Alternatively, the possibility cannot be completely ruled out that the
unipolar outflows are real and intrinsic phenomena in the process of
star formation.  Previous magnetohydrodynamics (MHD) simulations have
shown that unipolar outflows can indeed be driven by protostellar
systems formed during the gravitational collapse of magnetized
turbulent cloud cores \citep{2021A&A...656A..85M, 2024ApJ...963...20T,
  2024PASJ...76..823S}.  In particular, \citet{2024ApJ...963...20T}
showed that unipolar outflows can be formed in single low-mass
protostellar systems when the magnetic energy of the initial cloud
core is smaller than the turbulent energy of the initial cloud core.

While the previous simulations have reproduced unipolar outflows, the
conditions for the formation of unipolar outflows remain poorly
understood.  As mentioned above, \citet{2024ApJ...963...20T} suggested
a possible relation between unipolar outflows and the relative
strength of the turbulence to the magnetic field in the initial cloud
cores.  In their simulations, however, the turbulent energy of the
initial cloud core was fixed in the subsonic regime, with the sonic
Mach number of $\machs=0.86$.  This limitation raises the question of
whether unipolar outflows can also be formed under different turbulent
environments.  Exploring the range of the parameters of the turbulence
and magnetic field for the formation of unipolar outflows could
contribute to understanding the origin of the observed unipolar
outflows.

This study systematically investigates the formation of unipolar
outflows over a wide range of parameters by varying both the strengths
of the turbulence and magnetic field in the initial cloud cores, using
three-dimensional non-ideal MHD simulations.

The remainder of this paper is organized as follows.  Section
\ref{sec:Method_and_IC} describes the methods and initial conditions.
Section \ref{sec:results} presents the numerical results.  Section
\ref{sec:discussion} discusses the presented findings.  Finally,
section \ref{sec:summary} summarizes the results and findings.

\section{Methods and initial conditions}
\label{sec:Method_and_IC}

\subsection{Basic Equations and Numerical Method}
\label{subsec:numerical_method}

Our numerical method follows that described in
\citet{2024ApJ...963...20T}.  We solve the non-ideal MHD equations
with self-gravity:
\begin{align}
  \frac{D\vel}{Dt} &= -\frac{1}{\rho}
  \left\{
  \nabla\left(P+\frac{1}{2}|\magB|^2\right)-\nabla\cdot\left(\magB\magB\right)
  \right\}
  -\nabla\phi_{\rm{g}}, \label{eq:motion} \\
  \frac{D\magB}{Dt} &=
  \left(\magB\cdot\nabla\right)\vel-\magB\left(\nabla\cdot\vel\right) \nonumber \\
  &\ \ \ \  -\nabla \times \left\{
  \etaO (\nabla \times \magB)
  -\etaA ( (\nabla\times\magB)\times\magBhat ) \times \magBhat \right\}, \label{eq:induction} \\
  \nabla^2 \phi_{\rm{g}} &= 4 \pi G \rho, \label{eq:poisson}
\end{align}
where $\vel$ is the gas velocity, $\rho$ is the gas density, $P$ is
the gas pressure, $\magB$ is the magnetic field, $\phi_{\rm{g}}$ is
the gravitational potential, $G$ is the gravitational constant,
$\magBhat\equiv\magB/|\magB|$ is the unit vector along the magnetic
field, and $\etaO$ and $\etaA$ are the resistivities for the ohmic
dissipation and ambipolar diffusion of the non-ideal MHD effects.  The
Hall effect is ignored in this study.

We adopt the barotropic equation of state, which mimics the thermal
evolution of the cloud core calculated by \citet{2000ApJ...531..350M}:
\begin{equation}
  P = P(\rho) = c_{\rm{s,iso}}^2 \rho
  \left\{ 1+\left(\frac{\rho}{\rho_{\rm{crit}}}\right)^{\Gamma_{\rm{ad}}-1} \right\},
  \label{eq:eos}
\end{equation}
where $c_{\rm{s,iso}}=1.9\times10^4\cms$ is the isothermal sound speed
at the temperature of $10\thinspace\rm{K}$.
$\rho_{\rm{crit}}=4\times10^{-14}\gcm$ is the critical density at
which the thermal evolution changes from the isothermal to adiabatic.
We use the adiabatic index of $\Gamma_{\rm{ad}}=5/3$.

Equations (\ref{eq:motion})-(\ref{eq:eos}) are solved with the
smoothed particle hydrodynamics (SPH) method
\citep{1977AJ.....82.1013L, 1977MNRAS.181..375G, 1985A&A...149..135M}.
The ideal MHD part is solved using the Godunov smoothed particle
magnetohydrodynamics (GSPMHD) method \citep{2011MNRAS.418.1668I}.  The
hyperbolic divergence cleaning method \citep{2013ASPC..474..239I} is
used for the divergence-free condition.  The ohmic dissipation and
ambipolar diffusion are calculated following
\citet{2013MNRAS.434.2593T} and \citet{2014MNRAS.444.1104W}.  The
processes of the ohmic dissipation and ambipolar diffusion are
accelerated by the super-time-stepping (STS) method
\citep{Alexiades1996}, which is a numerical technique for reducing the
computational cost.  The self-gravity is calculated with the
Barnes-Hut octree algorithm with an opening angle of
$\theta_{\rm{gravity}}=0.5$ \citep{1986Natur.324..446B}.  The
numerical code is parallelized with the Message Passing Interface
(MPI).  The numerical code has already been applied to a variety of
problems (e.g., \cite{2011MNRAS.416..591T, 2013MNRAS.428.1321T,
  2013MNRAS.434.2593T, 2013MNRAS.436.1667T, 2015MNRAS.446.1175T,
  2015ApJ...810L..26T, 2015MNRAS.452..278T, 2017PASJ...69...95T,
  2018ApJ...868...22T, 2020MNRAS.492.5641T, 2020ApJ...896..158T,
  2021ApJ...913..148T, 2021ApJ...920L..35T, 2021PASJ...73L..25T,
  2023MNRAS.521.2661K, 2023PASJ...75..835T, 2024ApJ...963...20T}).

\subsection{Resistivity model}
\label{subsec:resistivity_model}

The simulations adopt the tabulated $\etaO$ and $\etaA$ of the
single-sized dust model with the dust size of $a_{\rm{d}}=0.035\mum$,
which is used in our previous study \citep{2024ApJ...963...20T}.  The
adopted resistivity tables are identical to those used in
\citet{2020ApJ...896..158T}; see Figure 1 of that paper for the
density and magnetic field dependence of $\etaO$ and $\etaA$.  The
resistivities are generated by the chemical reaction network
calculation \citep{2015ApJ...801...13S}.  The calculation includes the
ion species of $\rm{H^+}$, $\rm{H_2^+}$, $\rm{H_3^+}$, $\rm{HCO^+}$,
$\rm{Mg^+}$, $\rm{He^+}$, $\rm{C^+}$, $\rm{O^+}$, $\rm{O_2^+}$,
$\rm{H_3O^+}$, $\rm{OH^+}$, and $\rm{H_2O^+}$, and the neutral species
of $\rm{H}$, $\rm{H_2}$, $\rm{He}$, $\rm{CO}$, $\rm{O_2}$, $\rm{Mg}$,
$\rm{O}$, $\rm{C}$, $\rm{HCO}$, $\rm{H_2O}$, $\rm{OH}$, $\rm{N}$, and
$\rm{Fe}$.  The calculation also includes the neutral and singly
charged dust grains, $\rm{G^{0}}$, $\rm{G^{+}}$, and $\rm{G^{-}}$.
The calculation takes into account the cosmic-ray ionization,
gas-phase and dust-surface recombination, and ion-neutral reactions.
The indirect ionization by high-energy photons emitted by direct
cosmic-ray ionization (described as CRPHOT in the UMIST database) is
also considered in the calculation.  The initial abundance and
reaction rates are taken from the UMIST2012 database
\citep{2013A&A...550A..36M}.  The grain-ion and grain-grain collision
rates are calculated using the equations in
\citet{1987ApJ...320..803D}.  The chemical reaction network
calculation is conducted using the CVODE package \citep{Hindmarsh2005}
assuming the system is in the chemical equilibrium.  The resistivities
are calculated using the abundances of charged species in the
equilibrium state.  The momentum transfer rate between the charged and
neutral species is calculated using the equations in
\citet{2008A&A...484...17P}.  The temperature for the chemical
reaction network calculation is modeled as
$T_{\rm{chem}}=10(1+\gamma_{\rm{T}}(\rho/\rho_{\rm{crit}})^{(\gamma_{\rm{T}}-1)})
\thinspace\rm{K}$, where $\gamma_{\rm{T}}=7/5$.  We adopt
$\gamma_{\rm{T}}=7/5$ for consistency with our previous studies.
Although this value differs from $\Gamma_{\rm{ad}}=5/3$ adopted in the
barotropic equation of state, most of the computational domain,
particularly the outflow region, remains nearly isothermal throughout
the simulations.  Therefore, the influence of this difference is
expected to be minor.  It is assumed that the dust has the internal
density of $\rho_{\rm{d}}=2\gcm$ and the size of
$a_{\rm{d}}=0.035\mum$.  The dust-to-gas mass ratio is fixed to be
0.01 in the calculation.  The cosmic ray ionization rate is assumed to
be $\xi_{\rm{CR}}=10^{-17}\thinspace\rm{s^{-1}}$.

\subsection{Initial conditions}
\label{subsec:init_cond}

The simulations start from the gravitational collapse of isolated
molecular cloud cores including the turbulence and magnetic field
simultaneously.  The initial setup is the same as in
\citet{2024ApJ...963...20T}, except for the strengths of the
turbulence and magnetic field.

The initial cloud core follows the density profile of
\begin{align}
  \rho(r) &= \left\{
  \begin{array}{ll}
    f\rho_{\rm{c}}\varrho_{\rm{BE}}\left(r/a\right) & \rm{for}~{\it{r<R_{\rm{c}}}} \\
    f\rho_{\rm{c}}\varrho_{\rm{BE}}\left(R_{\rm{c}}/a\right)
    \left(\frac{r}{R_{\rm{c}}}\right)^{-4} & \rm{for}~{\it{R_{\rm{c}} \leq r<{\rm 10}R_{\rm{c}}}}
  \end{array}
  \right.,
  \label{eq:prof_rho} \\
  a &\equiv c_{\rm{s,iso}} \left( \frac{1}{4\pi G \rho_{\rm{c}}} \right)^{1/2},
  \label{eq:prof_a}
\end{align}
where $\rho_{\rm{c}}$ is the characteristic density,
$R_{\rm{c}}=6.45a$ is the radius of the initial cloud core, and
$\varrho_{\rm{BE}}$ is the non-dimensional density profile of the
Bonnor-Ebert sphere \citep{1956MNRAS.116..351B, 1955ZA.....37..217E}.
The density enhancement factor $f$ is written as $f=0.84/\alpha$ (see
the appendix \ref{app:sec:pq_core}), where $\alpha$ is defined as
\begin{equation}
  \alpha = \frac{E_{\rm{thm}}}{|E_{\rm{grav}}|},
  \label{eq:alpha}
\end{equation}
with $E_{\rm{thm}}$ and $E_{\rm{grav}}$ being the thermal and
gravitational energies of the initial cloud core of $r \leq
R_{\rm{c}}$.  The thermal energy $E_{\rm{thm}}$ can be written as
\begin{equation}
  E_{\rm{thm}} = \frac{3}{2} M_{\rm{c}} c_{\rm{s,iso}}^2,
  \label{eq:Ethm}
\end{equation}
where $M_{\rm{c}}$ is the mass of the initial cloud core within the
core radius $R_{\rm{c}}$, defined as $M_{\rm{c}}=M(R_{\rm{c}})$.  The
enclosed mass $M(r)$ is given by equation (\ref{app:eq:Menclose}).

The initial cloud core has the temperature of
$T_{\rm{c}}=10\thinspace\rm{K}$, the radius of
$R_{\rm{c}}=4.8\times10^{3}\au$, the mass of $M_{\rm{c}}=1\msun$
within the radius $R_{\rm{c}}$ ($\sim2.1\msun$ in the entire domain of
$r<10R_{\rm{c}}$), and $\alpha=0.4$, which are determined by
specifying $f=2.1$ and the central density of
$\rho_{0}=f\rho_{\rm{c}}=7.3\times10^{-18}\gcm$.  The corresponding
free-fall time is calculated as
$\tff=(3\pi/(32G\rho_{0}))^{1/2}=2.5\times10^{4}\yr$.  The simulations
resolve $M_{\rm{c}}=1\msun$ with the number of SPH particles of
$N_{\rm{SPH}}=10^6$, which corresponds to the mass resolution of
$M_{\rm{c}}/N_{\rm{SPH}}=10^{-6}\msun$.  The number of all the SPH
particles in the entire domain of $r<10R_{\rm{c}}$ is $\sim
2.1\msun/10^{-6}\msun=2.1\times10^{6}$.

The initial cloud cores are characterized by the strengths of the
turbulence and magnetic field.  The initial cloud core has the
divergence-free turbulent velocity field with the velocity power
spectrum of $P_{v}(k) \propto k^{-4}$, where $k$ is the wavenumber.
Such a velocity field reproduces the observed rotational properties of
molecular cloud cores \citep{2000ApJ...543..822B} and is adopted in
\citet{2024ApJ...963...20T}.  The strength of the turbulence is
described by the sonic Mach number:
\begin{equation}
  \machs = \frac{\sqrt{3}\sigma_{\rm{v}}}{c_{\rm{s,iso}}},
  \label{eq:Ms}
\end{equation}
where $\sigma_{\rm{v}}$ is the one-dimensional velocity dispersion:
\begin{equation}
  \sigma_{\rm{v}} = \sqrt{\frac{2E_{\rm{turb}}}{3M_{\rm{c}}}}.
  \label{eq:1D_Vdisp}
\end{equation}
Equation (\ref{eq:Ms}) is rewritten as the ratio of the turbulent to
gravitational energies:
\begin{equation}
  \gturb = \frac{E_{\rm{turb}}}{|E_{\rm{grav}}|},
  \label{eq:gturb}
\end{equation}
where $E_{\rm{turb}}$ is the turbulent energy of the initial cloud
core of $r \leq R_{\rm{c}}$.  The initial cloud core has a
non-vanishing net angular momentum of $|\Jcore|$ due to the stochastic
nature of the turbulence.  The direction of $\Jcore$ is chosen as the
$z$-axis of the simulation.

The turbulent velocity field is generated following
\citet{2013MNRAS.428.1321T}, which is also used in our previous
studies \citep{2020MNRAS.492.5641T, 2021PASJ...73L..25T,
  2024ApJ...963...20T}.  All the models use the same seed for the
turbulent velocity field.  The turbulent velocity field is assigned to
the cloud core only for $r<R_{\rm{c}}$, and is set to zero for
$R_{\rm{c}}\leq r< 10R_{\rm{c}}$.  Note that the initial velocity
field consists only of the turbulent velocity field, and it is not the
superposition of the turbulent and coherent rotational velocity
fields.  Appendix \ref{app:sec:numerical_convergence_test} presents an
additional simulation with a different random seed.  The qualitative
outflow morphology remains unchanged in this test.

The initial cloud core has the axisymmetric magnetic field modeled by
\citet{2020ApJ...896..158T}.  The magnetic field is oriented
approximately along the $z$-axis and is parallel to the net angular
momentum vector of the initial cloud core, $\Jcore$.  The strength of
the magnetic field is conventionally expressed by the dimensionless
mass-to-flux ratio:
\begin{eqnarray}
  \magmu &=&
  \left(\frac{M_{\rm{c}}}{\Phi_{\rm{mag}}\left(R_{\rm{c}}\right)}\right) \left/
  \left(\frac{M_{\rm{c}}}{\Phi_{\rm{mag}}}\right)_{\rm{crit}} \right. ,
  \label{eq:mass-to-flux}
\end{eqnarray}
where $\Phi_{\rm{mag}}\left(R_{\rm{c}}\right)$ is the magnetic flux of
the initial cloud core.
$(M_{\rm{c}}/\Phi_{\rm{mag}})_{\rm{crit}}=(0.53/3\pi)(5/G)^{1/2}$ is
the critical mass-to-flux ratio on stability for uniform spheres
\citep{1976ApJ...210..326M}.  The dimensionless mass-to-flux ratio
with the constant value of the central magnetic field $B_{\rm{c}}$ is
calculated as
\begin{eqnarray}
  \magmu_{\rm{const}} =
  \left( \frac{M_{\rm{c}}}{\pi R_{\rm{c}}^2B_{\rm{c}}} \right) \left/
  \left( \frac{M_{\rm{c}}}{\Phi_{\rm{mag}}} \right)_{\rm{crit}} \right.
  = \frac{\magmu}{2}.
  \label{eq:mu_const}
\end{eqnarray}
We note that $\magmu_{\rm{const}}$ would be a suitable indicator to
compare the strength of the magnetic field of this study with those of
previous studies because we focus on the time evolution of the central
region of the cloud core (see \cite{2020ApJ...896..158T} for details).
Equation (\ref{eq:mass-to-flux}) is rewritten as the ratio of the
magnetic to gravitational energies:
\begin{eqnarray}
  \gmag = \frac{E_{\rm{mag}}}{|E_{\rm{grav}}|}, \label{eq:gmag}
\end{eqnarray}
where $E_{\rm{mag}}$ is the magnetic energy of the initial cloud core
without the surrounding medium of $\rho(r)\propto r^{-4}$.  The plasma
beta parameter of the initial cloud core can be estimated as
$\overline{\beta}_{\rm{plasma}}\sim2\bthm/3$, where $\bthm$ is the
ratio of the thermal to magnetic energies calculated as
$\bthm=E_{\rm{thm}}/E_{\rm{mag}}=\alpha/\gmag$
\citep{2017ApJ...845...32K}.

All the simulations are conducted to $t=\tpform+\tpend\gtrsim 2\tff$,
where $\tpform$ is the time at which the central density becomes
higher than $1.0\times10^{-11}\gcm$.  The elapsed time after the
protostar formation is defined as $\tp \equiv t-\tpform$, and $\tp$ at
the end of the simulation is denoted as $\tpend$.

A total of 24 models were conducted with the different sets of values
for $\machs$ and $\magmu$.  The model names and parameters are
summarized in Table \ref{table:summary_parameters}.

\begin{table*}
  \centering
  \tbl{Parameters for 24 simulation models.\footnotemark[$*$]}{%
    \begin{tabular}{ccccccccccccc}
      \hline
      \hline
      Model &  
      $\machs$ &  
      $\magmu$ &  
      $\gturb$ &  
      $\gmag$ &  
      $|\Jcore|\thinspace[\rm{g\thinspace cm^2\thinspace s^{-1}}]$ &  
      $B_{\rm{c}}\thinspace[\mu\rm{G}]$ &  
      $\magmu_{\rm{const}}$ &  
      $\tpform\thinspace[\rm{yr}]$ &  
      $\tpend\thinspace[\rm{yr}]$ &  
      $\machA$ &  
      Outflow \\ 
      \hline
      \hline
      M027MF2  & 0.27 & 2  & $1.0\times10^{-2}$ & $4.2\times10^{-1}$ & $1.4\times10^{53}$ & $2.5\times10^2$ & 1 & $8.7\times10^{4}$  & $3.4\times10^{3}$ & 0.15 & No           \\ 
      M027MF4  & 0.27 & 4  & $1.0\times10^{-2}$ & $1.0\times10^{-1}$ & $1.4\times10^{53}$ & $1.3\times10^2$ & 2 & $4.7\times10^{4}$  & $5.4\times10^{3}$ & 0.31 & Bipolar      \\ 
      M027MF8  & 0.27 & 8  & $1.0\times10^{-2}$ & $2.6\times10^{-2}$ & $1.4\times10^{53}$ & $6.3\times10^1$ & 4 & $4.3\times10^{4}$  & $3.4\times10^{3}$ & 0.62 & Bipolar(S)   \\ 
      M027MF16 & 0.27 & 16 & $1.0\times10^{-2}$ & $6.5\times10^{-3}$ & $1.4\times10^{53}$ & $3.1\times10^1$ & 8 & $4.2\times10^{4}$  & $1.5\times10^{4}$ & 1.2  & Unipolar(S)  \\ 
      \hline
      M043MF2  & 0.43 & 2  & $2.5\times10^{-2}$ & $4.2\times10^{-1}$ & $2.2\times10^{53}$ & $2.5\times10^2$ & 1 & $9.3\times10^{4}$  & $5.1\times10^{3}$ & 0.25 & No           \\ 
      M043MF4  & 0.43 & 4  & $2.5\times10^{-2}$ & $1.0\times10^{-1}$ & $2.2\times10^{53}$ & $1.3\times10^2$ & 2 & $4.9\times10^{4}$  & $7.3\times10^{3}$ & 0.49 & Bipolar      \\ 
      M043MF8  & 0.43 & 8  & $2.5\times10^{-2}$ & $2.6\times10^{-2}$ & $2.2\times10^{53}$ & $6.3\times10^1$ & 4 & $4.4\times10^{4}$  & $6.5\times10^{3}$ & 0.98 & Bipolar(S)   \\ 
      M043MF16 & 0.43 & 16 & $2.5\times10^{-2}$ & $6.5\times10^{-3}$ & $2.2\times10^{53}$ & $3.1\times10^1$ & 8 & $4.4\times10^{4}$  & $1.5\times10^{4}$ & 2.0  & Unipolar(S)  \\ 
      \hline
      M061MF2  & 0.61 & 2  & $5.0\times10^{-2}$ & $4.2\times10^{-1}$ & $3.1\times10^{53}$ & $2.5\times10^2$ & 1 & $1.0\times10^{5}$  & $7.0\times10^{3}$ & 0.35 & No           \\ 
      M061MF4  & 0.61 & 4  & $5.0\times10^{-2}$ & $1.0\times10^{-1}$ & $3.1\times10^{53}$ & $1.3\times10^2$ & 2 & $5.1\times10^{4}$  & $7.6\times10^{3}$ & 0.69 & Bipolar      \\ 
      M061MF8  & 0.61 & 8  & $5.0\times10^{-2}$ & $2.6\times10^{-2}$ & $3.1\times10^{53}$ & $6.3\times10^1$ & 4 & $4.6\times10^{4}$  & $1.5\times10^{4}$ & 1.4  & Unipolar     \\ 
      M061MF16 & 0.61 & 16 & $5.0\times10^{-2}$ & $6.5\times10^{-3}$ & $3.1\times10^{53}$ & $3.1\times10^1$ & 8 & $4.6\times10^{4}$  & $1.3\times10^{4}$ & 2.8  & Unipolar(S)  \\ 
      \hline
      M086MF2  & 0.86 & 2  & $1.0\times10^{-1}$ & $4.2\times10^{-1}$ & $4.4\times10^{53}$ & $2.5\times10^2$ & 1 & $1.4\times10^{5}$  & $1.5\times10^{4}$ & 0.49 & No           \\ 
      M086MF4  & 0.86 & 4  & $1.0\times10^{-1}$ & $1.0\times10^{-1}$ & $4.4\times10^{53}$ & $1.3\times10^2$ & 2 & $5.5\times10^{4}$  & $1.1\times10^{4}$ & 0.98 & Bipolar      \\ 
      M086MF8  & 0.86 & 8  & $1.0\times10^{-1}$ & $2.6\times10^{-2}$ & $4.4\times10^{53}$ & $6.3\times10^1$ & 4 & $4.9\times10^{4}$  & $2.3\times10^{4}$ & 2.0  & Unipolar     \\ 
      M086MF16 & 0.86 & 16 & $1.0\times10^{-1}$ & $6.5\times10^{-3}$ & $4.4\times10^{53}$ & $3.1\times10^1$ & 8 & $4.8\times10^{4}$  & $2.3\times10^{4}$ & 3.9  & Unipolar     \\ 
      \hline
      M12MF2   & 1.2  & 2  & $2.0\times10^{-1}$ & $4.2\times10^{-1}$ & $6.2\times10^{53}$ & $2.5\times10^2$ & 1 & $\rm{No~protstar}$ &                   & 0.69 & No           \\ 
      M12MF4   & 1.2  & 4  & $2.0\times10^{-1}$ & $1.0\times10^{-1}$ & $6.2\times10^{53}$ & $1.3\times10^2$ & 2 & $6.3\times10^{4}$  & $1.6\times10^{3}$ & 1.4  & Unipolar(SF) \\ 
      M12MF8   & 1.2  & 8  & $2.0\times10^{-1}$ & $2.6\times10^{-2}$ & $6.2\times10^{53}$ & $6.3\times10^1$ & 4 & $5.5\times10^{4}$  & $2.1\times10^{4}$ & 2.8  & Unipolar     \\ 
      M12MF16  & 1.2  & 16 & $2.0\times10^{-1}$ & $6.5\times10^{-3}$ & $6.2\times10^{53}$ & $3.1\times10^1$ & 8 & $5.4\times10^{4}$  & $2.5\times10^{4}$ & 5.5  & Unipolar     \\ 
      \hline
      M17MF2   & 1.7  & 2  & $4.0\times10^{-1}$ & $4.2\times10^{-1}$ & $8.8\times10^{53}$ & $2.5\times10^2$ & 1 & $\rm{No~protstar}$ &                   & 0.98 & No           \\ 
      M17MF4   & 1.7  & 4  & $4.0\times10^{-1}$ & $1.0\times10^{-1}$ & $8.8\times10^{53}$ & $1.3\times10^2$ & 2 & $8.8\times10^{4}$  & $1.1\times10^{4}$ & 2.0  & Unipolar(SF) \\ 
      M17MF8   & 1.7  & 8  & $4.0\times10^{-1}$ & $2.6\times10^{-2}$ & $8.8\times10^{53}$ & $6.3\times10^1$ & 4 & $7.4\times10^{4}$  & $1.9\times10^{4}$ & 3.9  & Unipolar     \\ 
      M17MF16  & 1.7  & 16 & $4.0\times10^{-1}$ & $6.5\times10^{-3}$ & $8.8\times10^{53}$ & $3.1\times10^1$ & 8 & $7.1\times10^{4}$  & $3.9\times10^{4}$ & 7.8  & Unipolar     \\ 
      \hline
      \hline
  \end{tabular}}\label{table:summary_parameters}
  \begin{tabnote}
    \footnotemark[$*$]
    Model name,
    sonic Mach number $\machs$,
    dimensionless mass-to-flux ratio $\magmu$,
    ratio of the turbulent to gravitational energies $\gturb=E_{\rm{turb}}/|E_{\rm{grav}}|$,
    ratio of the magnetic to gravitational energies $\gmag=E_{\rm{mag}}/|E_{\rm{grav}}|$,
    total angular momentum $|\Jcore|$,
    strength of the central magnetic field $B_{\rm{c}}$,
    and
    dimensionless mass-to-flux ratio with the constant value of the central magnetic field $\magmu_{\rm{const}}$.
    $\tpform$ is the time at which the protostar forms.
    $\tpend$ is the elapsed time after the protostar formation at the end of the simulation.
    $\machA$ is the Alfv\'{e}n Mach number of the initial cloud core.
    The last column indicates the morphology of the driven outflow.
    Capitals ``S'' and ``F'' mean the small ($\lesssim5\times10^{2}\au$) and failed outflows.
  \end{tabnote}
\end{table*}

\section{Results}
\label{sec:results}

\subsection{Overview: formation of unipolar and bipolar outflows}
\label{subsec:overview}

\begin{figure*}
  \begin{center}
    \includegraphics[clip,width=0.85\linewidth]{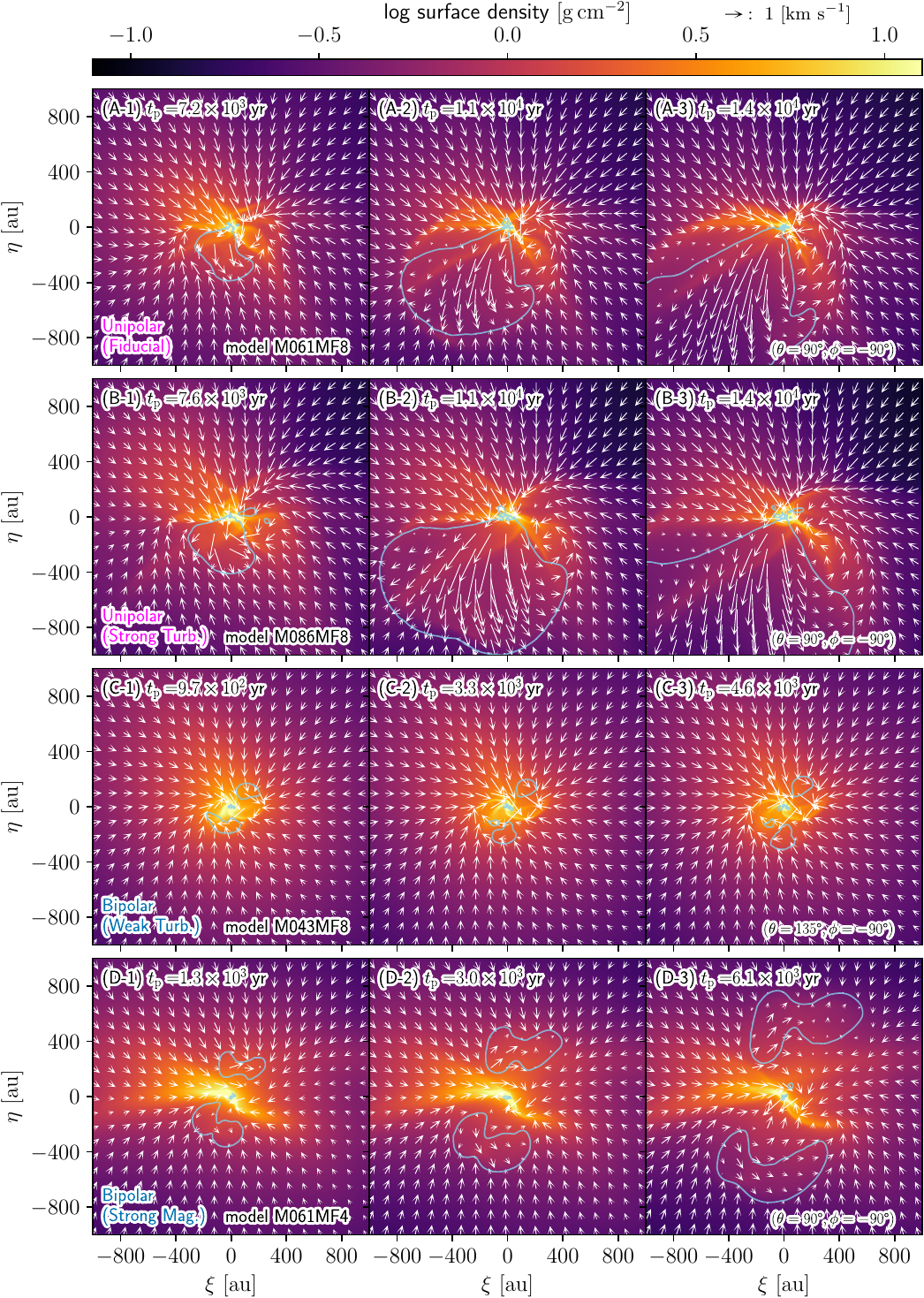}
  \end{center}
  \caption{ Evolution of the surface density on the $\xi-\eta$ plane
    for the models M061MF8 ($\machs=0.61$, $\magmu=8$), M086MF8
    ($\machs=0.86$, $\magmu=8$), M043MF8 ($\machs=0.43$, $\magmu=8$),
    and M061MF4 ($\machs=0.61$, $\magmu=4$).  The vertical and
    horizontal axes are chosen along the unit vectors
    $\bm{e}_{\eta}=(\bm{e}_{z}-(\bm{e}_{z}\cdot\nhat)\nhat)/
    |\bm{e}_{z}-(\bm{e}_{z}\cdot\nhat)\nhat|$ and
    $\bm{e}_{\xi}=\bm{e}_{\eta}\times\nhat$, where $\bm{e}_{z}$ is the
    unit vector in the direction of the $z$-axis and $\nhat$ is the
    normal vector written in spherical coordinates
    $(r,\thinspace\theta,\thinspace\phi)$ by
    $\nhat=(\sin\theta\cos\phi,\thinspace\sin\theta\sin\phi,\thinspace\cos\theta)$
    (see Fig. 2 in \cite{2011PASJ...63..147T} for more details).
    Thus, $-\nhat$ represents the line-of-sight direction from the
    observer to the protostellar system.  To calculate the surface
    density, first, the origin of a coordinate system is shifted to
    the center of mass of the system.  Then, $\bm{e}_{\xi}$ and
    $\bm{e}_{\eta}$ are calculated.  Finally the integrations of the
    surface density, $\Sigma=\int\rho ds$, are conducted along $\nhat$
    in a range from $s=-2R_{\rm{c}}$ to $s=2R_{\rm{c}}$.  The
    cut-plane velocity at $s=0$ is plotted as the white arrows.  The
    reference arrow plotted on the top right corresponds to
    $1\thinspace\kms$.  Blue solid lines show the velocity contours at
    the radial velocity of $v_r=0$.  \\
    {Alt text: Surface density evolution of outflows for four models.}
  }
  \label{fig:surface_density_map}
\end{figure*}

First, we present the evolution of the surface density for four
typical models in which the outflow is formed, as shown in Figure
\ref{fig:surface_density_map}: (A) M061MF8 (fiducial), (B) M086MF8
(with strong turbulence), (C) M043MF8 (with weak turbulence), and (D)
M061MF4 (with strong magnetic field).  Blue solid lines approximately
outline the outflow region.

The unipolar outflow is formed in panels (A) and (B).  From (A-1) to
(A-3), the unipolar outflow is driven from the protostellar system at
a few $\rm{km}\thinspace\rm{s}^{-1}$ and grows gradually.  In panels
(B), the unipolar outflow exhibits a similar growth pattern to that in
panels (A).

In contrast to panels (A) and (B), the bipolar outflow is formed in
panels (C) and (D).  The results in panels (A)-(D) suggest that the
morphologies of the outflow vary depending on the strengths of the
turbulence ($\machs$) and magnetic field ($\magmu$).

The opening angle of the unipolar outflow in the panels (B) is larger
than in panels (A).  This suggests that the strength of the turbulence
regulates the opening angle of the unipolar outflow.  The size of the
bipolar outflow is different between panels (C) and (D).  The bipolar
outflow in panels (C) is smaller, with the size of less than
$\sim5\times10^{2}\au$.  The bipolar outflow in panels (D) grows
gradually and extends to $\sim10^{3}\au$.  The arc-like structures of
the infalling envelope are naturally formed in all the panels.

The dependence of the obtained outflow morphologies on the numerical
resolution is examined in Appendix
\ref{app:sec:numerical_convergence_test}.  The qualitative outflow
morphology remains unchanged in these tests.

\subsection{Morphologies of outflows}
\label{subsec:morphologies}

\begin{figure*}
  \begin{center}
    \includegraphics[clip,width=0.95\linewidth]{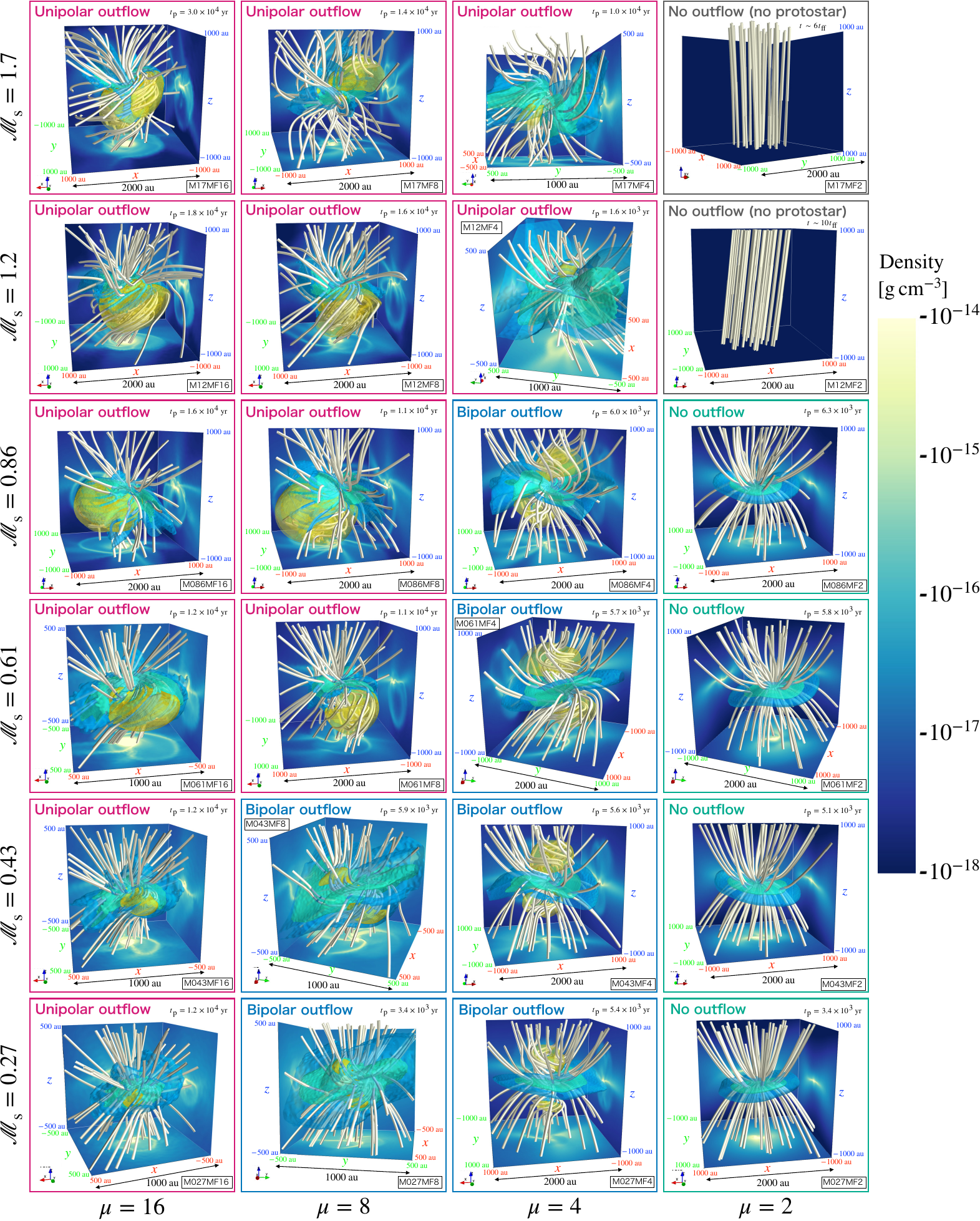}
  \end{center}
  \caption{ Three-dimensional views of the density distribution and
    magnetic field structure for all the models.  Yellow isosurfaces
    show the density of $3.2\times10^{-18}\gcm$, $10^{-17}\gcm$,
    $3.2\times10^{-17}\gcm$, and $10^{-16}\gcm$ with the radial
    velocity of $v_r>0$, representing the outflows.  White lines show
    the magnetic field lines.  Blue and green isosurfaces show the
    density of $3.2\times10^{-17}\gcm$ and $10^{-16}\gcm$ with
    $v_r<0$, representing the infalling envelopes.  Cut-plane
    densities on $x-y$ ($z=0$), $x-z$ ($y=0$), and $y-z$ ($x=0$)
    planes are projected for each panel.  The origin of a coordinate
    system is shifted to the center of mass of the system.  \\
    {Alt text: 3D density and magnetic field structure.}
  }
  \label{fig:3Dview}
\end{figure*}

Figure \ref{fig:3Dview} shows the three-dimensional views of the
density distribution and magnetic field structure for all the models
when the outflow is most clearly characterized.  The unipolar outflows
are formed in a wide range of the initial parameters $\machs$ and
$\magmu$.  Specifically, they are formed in the models M027MF16,
M043MF16, M061MF8, M061MF16, M086MF8, M086MF16, M12MF4, M12MF8,
M12MF16, M17MF4, M17MF8, and M17MF16 (magenta boxes in Figure
\ref{fig:3Dview}).  These models have the large values of $\machs$ and
$\magmu$.  This indicates that the unipolar outflows are formed in the
cloud cores with the strong turbulence (large $\machs$) and weak
magnetic field (large $\magmu$).

The direction of the unipolar outflow varies across the simulations.
The opening angle of the unipolar outflow tends to increase with
larger $\machs$ and $\magmu$.  The unipolar outflow remains small in
size, staying less than $\sim5\times10^{2}\au$ in the models M027MF16,
M043MF16, M061MF16, M12MF4, and M17MF4 until the end of the
simulation.

The bipolar outflows are formed in the models M027MF4, M027MF8,
M043MF4, M043MF8, M061MF4, and M086MF4 (blue boxes in Figure
\ref{fig:3Dview}).  The asymmetrical features of the bipolar outflow,
such as the bending and different size of each component, naturally
emerge because of the turbulent accretion.  The bipolar outflow
remains small in size, staying less than $\sim5\times10^{2}\au$ in the
models M027MF8 and M043MF8 until the end of the simulation.

The outflows are not formed in the models with the strong magnetic
field of $\magmu=2$.  In the subsonic models M027MF2, M043MF2,
M061MF2, and M086MF2 (green boxes in Figure \ref{fig:3Dview}), the
infalling envelope shapes the disk-shaped flattened structure, which
is so-called the pseudo-disk \citep{1993ApJ...417..220G,
  1993ApJ...417..243G}.  In the supersonic models M12MF2 and M17MF2
(gray boxes in Figure \ref{fig:3Dview}), the cloud cores are
dissipated by the strong turbulence, and protostars are not formed.

\begin{figure}
  \begin{center}
    \includegraphics[clip,width=0.95\linewidth]{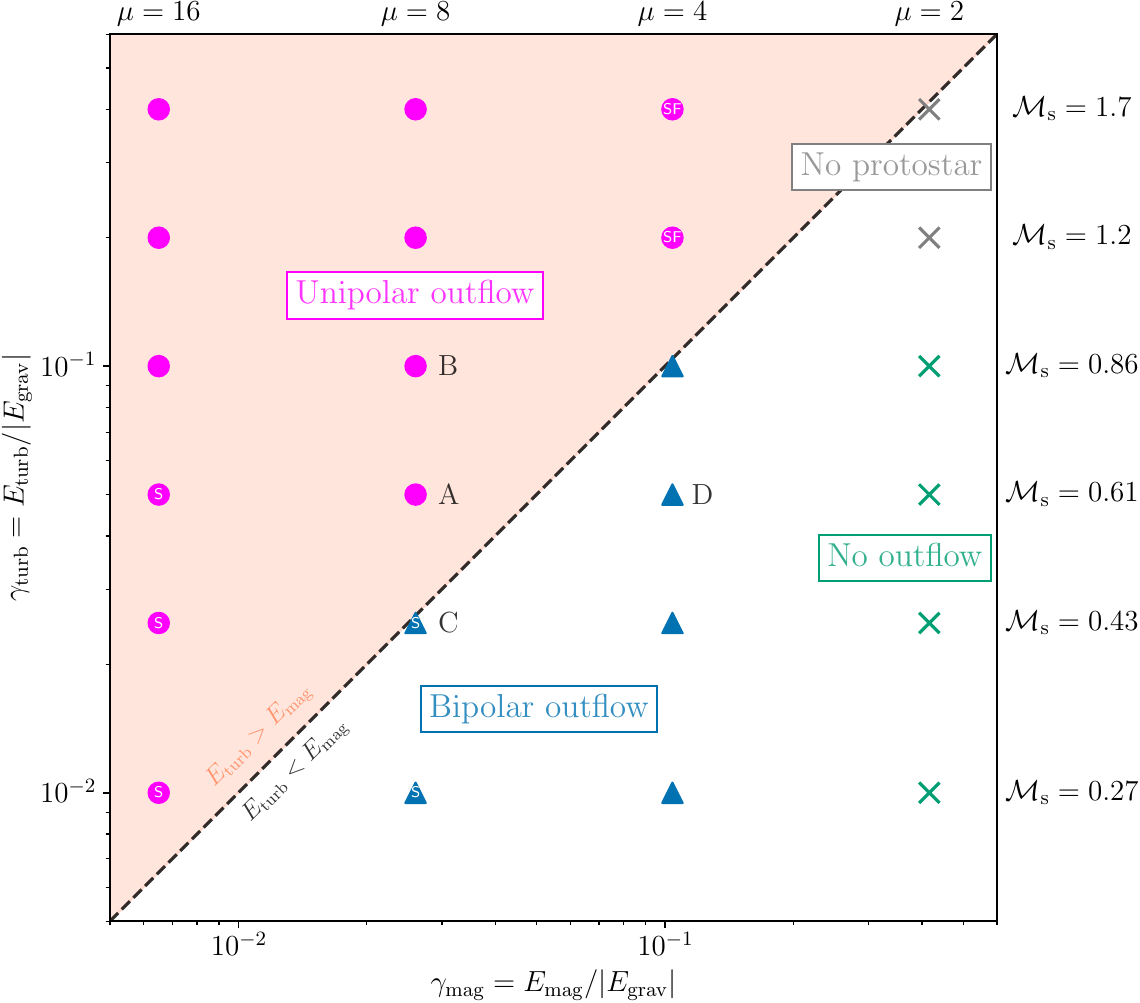}
  \end{center}
  \caption{ Outflow morphologies plotted on $(\gmag,\thinspace\gturb)$
    plane, with the order corresponding to Figure \ref{fig:3Dview}.
    The circles (magenta), triangles (blue), and crosses (green and
    gray) indicate the unipolar, bipolar, and no outflows.  Dashed
    line means $\gturb=\gmag$ ($E_{\rm{turb}}=E_{\rm{mag}}$).  Shaded
    region shows $\gturb>\gmag$ ($E_{\rm{turb}}>E_{\rm{mag}}$)
    Capitals ``S'' and ``F'' mean the small
    ($\lesssim5\times10^{2}\au$) and failed outflows.  \\
    {Alt text: Scatter plot of simulation results on
      $(\gmag,\thinspace\gturb)$ plane.}
  }
  \label{fig:outflow_morphologies_plane_E}
\end{figure}

To understand the relation between the initial parameters and the
outflow morphologies more quantitatively, we plot the outflow
morphologies on the $\gmag-\gturb$ plane as in Figure
\ref{fig:outflow_morphologies_plane_E}.  This figure clearly shows
that the unipolar outflows are formed in the region of $\gturb>\gmag$
(i.e., $E_{\rm{turb}}>E_{\rm{mag}}$).  This suggests that the ratio of
the turbulent to magnetic energies of the cloud core plays a crucial
role in the formation of unipolar outflows.

Several of the driven outflows remain small in size
($\lesssim5\times10^{2}\au$) and/or fail to grow during the evolution,
which are indicated by the capitals ``S'' and ``F'' in Figure
\ref{fig:outflow_morphologies_plane_E}.  This is because that the ram
pressure caused by the strong turbulent accretion suppresses the
outflow growth.  While the small outflows are maintained until the end
of the simulations, the failed outflows are finally collapsed and
dissipated before the end of the simulations.

\subsection{Evolution of outflow physical quantities}
\label{subsec:time_evol_outflow}

\begin{figure}
  \begin{center}
    \includegraphics[clip,width=0.95\linewidth]{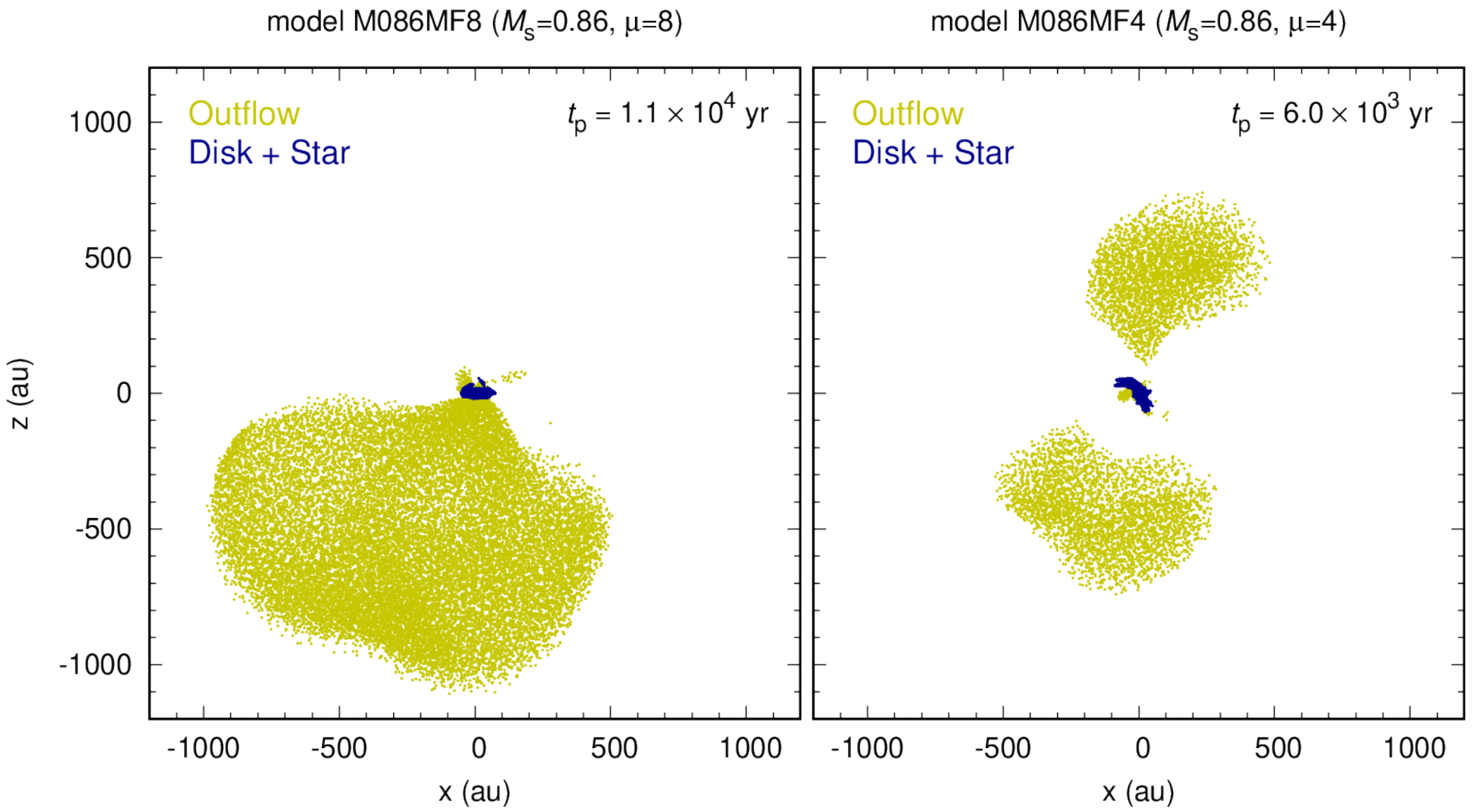}
  \end{center}
  \caption{ Distribution of the outflow particles (yellow dots)
    projected from the three-dimensional space onto the $x-z$ plane.
    Typical unipolar (left) and bipolar (right) outflows are plotted,
    which correspond to the models M086MF4 and M086MF8 in Figure
    \ref{fig:3Dview}.  Blue dots show the particles that satisfy
    $\rho>\times 10^{-14}\gcm$, corresponding to the protoplanetary
    disk and protostar.  \\
    {Alt text: Scatter plot of outflow particles.}
  }
  \label{fig:example_outflow_particles}
\end{figure}
    
As shown in section \ref{subsec:morphologies}, the outflow
morphologies are different in each simulation.  To investigate the
difference more quantitatively, we calculate the time evolution of the
physical quantities of the outflow.  The outflow is defined as a
region that satisfies the following conditions:
\begin{align}
  &v_{r}=\frac{(\vel-\vel_{\rm{p}})\cdot(\rad-\rad_{\rm{p}})}{|\rad-\rad_{\rm{p}}|}>0,
  \label{eq:cond:outflow:vr} \\
  &\rho_{\rm{out,min}}<\rho<\rho_{\rm{out,max}},
  \label{eq:cond:outflow:rho} \\
  &|\rad-\rad_{\rm{p}}|<L_{\rm{out,max}},
  \label{eq:cond:outflow:radius}
\end{align}
where $\rad_{\rm{p}}$ and $\vel_{\rm{p}}$ are the position and
velocity of the center of mass of the system, which correspond to the
position and velocity of the protostar.
$\rho_{\rm{out,min}}=3.2\times10^{-18}\gcm$ and
$\rho_{\rm{out,max}}=10^{-14}\gcm$ are adopted in this analysis.
$L_{\rm{out,max}}$ is the size of the outflow.

As reported by \citet{2024ApJ...963...20T}, protostellar systems with
the unipolar outflow are accelerated by the linear momentum transport
from the unipolar outflow and move outwards from the central region of
the parent cloud core.  In addition, the turbulence stochastically
perturbs the velocity of the gas.  As a result, some of the SPH
particles appear to be outflowing from some of the regions, especially
around $\simeq R_{\rm{c}}$, even though they are not driven by the
magnetic effects.  To exclude such pretender particles, we first
determine $L_{\rm{out,max}}$ as the maximum distance of the regions
satisfying the three conditions (i)
$|\rad-\rad_{\rm{p}}|<5\times10^3\au$, (ii) $\beta_{\rm{plasma}}<0.5$,
and (iii) inequalities (\ref{eq:cond:outflow:vr}) and
(\ref{eq:cond:outflow:rho}), where $\beta_{\rm{plasma}}$ is the plasma
beta parameter of the gas.  We then select the outflow particles based
on inequalities (\ref{eq:cond:outflow:vr}),
(\ref{eq:cond:outflow:rho}), and (\ref{eq:cond:outflow:radius}).
Figure \ref{fig:example_outflow_particles} shows examples of the
selected outflow particles for the unipolar and bipolar outflows based
on the above procedure.  We have confirmed that our procedure excludes
the pretender outflow particles.

\begin{figure}
  \begin{center}
    \includegraphics[clip,width=0.95\linewidth]{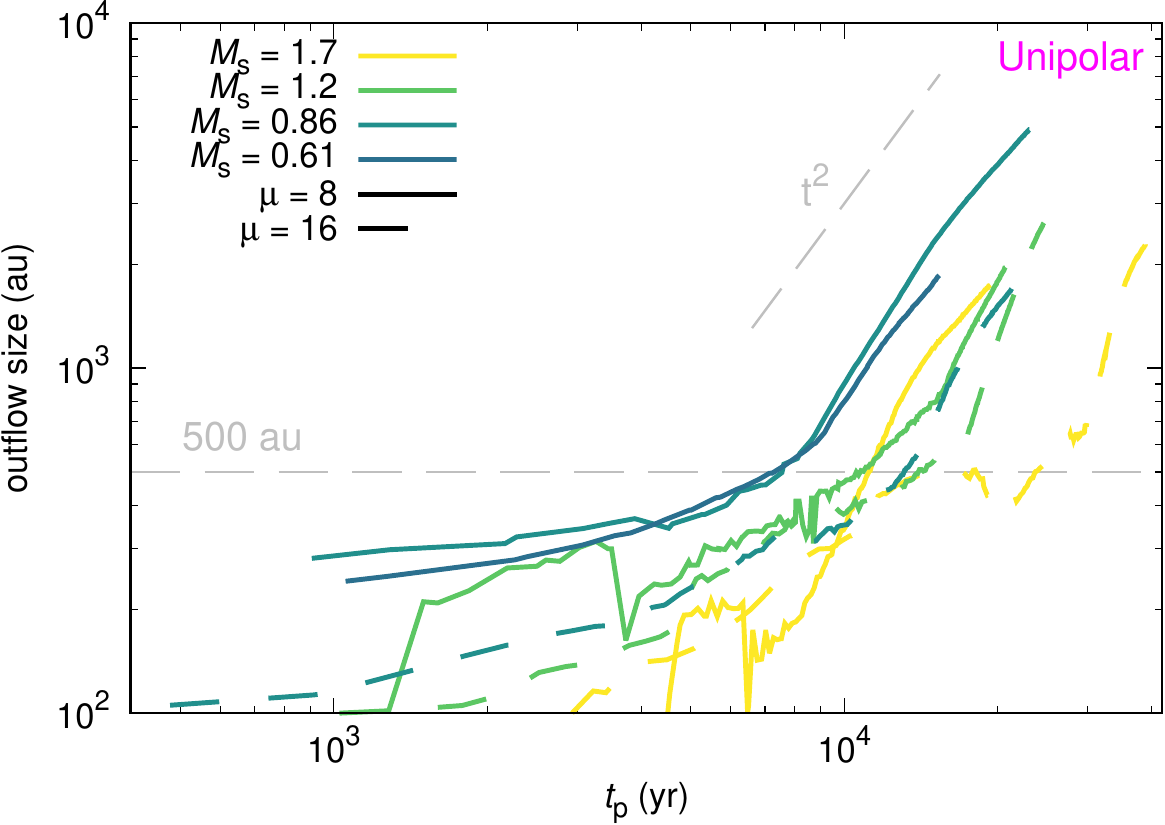}
    \includegraphics[clip,width=0.95\linewidth]{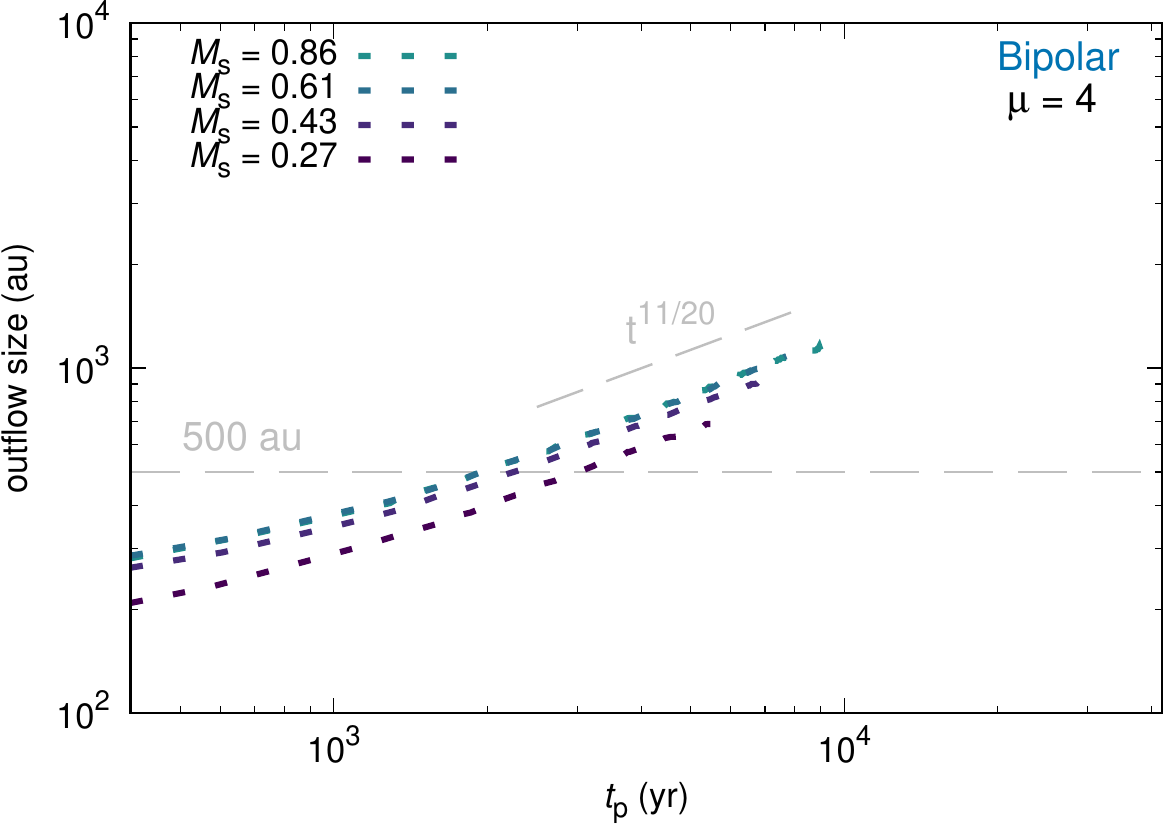}
  \end{center}
  \caption{ Time evolution of the size of the unipolar (upper) and
    bipolar (lower) outflows.  The dotted, solid, and dashed lines
    correspond to the models with $\magmu=4$, $\magmu=8$, and
    $\magmu=16$.  The difference in the colors corresponds to that in
    $\machs$.  The models M17MF8 (yellow solid), M12MF8 (green solid),
    M086MF8 (blue green solid), M061MF8 (blue solid), M17MF16 (yellow
    dashed) M12MF16 (green dashed), and M086MF16 (blue green dashed)
    are selected as the unipolar case.  The models M086MF4 (blue green
    dotted), M061MF4 (blue dotted), M043MF4 (purple dotted), and
    M027MF4 (dark purple dotted) are selected as the bipolar case.
    The small and failed outflows are excluded in the plots.  \\
    {Alt text: Line graphs showing the evolution of sizes of unipolar
      and bipolar outflows.}
  }
  \label{fig:time_evol_outflow_size}
\end{figure}

Figure \ref{fig:time_evol_outflow_size} shows the time evolution of
the size of the unipolar and bipolar outflows.  In the upper panel,
the size of the unipolar outflows gradually increases to
$\sim5\times10^{2}\au$, and then reaches $\sim2\times10^{3}\au$ after
$\tp\sim10^{4}\yr$.  The size of the unipolar outflows is roughly
proportional to $\tp^{2}$, indicating that the size increases with
nearly constant acceleration.  This is because that the protostellar
system is accelerated by the linear momentum transport from the
unipolar outflow, which is referred to as the \textit{protostellar
  rocket effect} in \citet{2024ApJ...963...20T}.

Here, the protostellar rocket effect is a consequence of the linear
momentum transport by the unipolar outflow rather than the driver of
the outflow itself.  We also note that the outflow size is measured
relative to the position of the protostellar system.  Therefore, the
measured outflow size reflects not only the propagation of the outflow
itself but also the motion of the protostellar system.

In the lower panel, the size of the bipolar outflows increases
monotonically in the evolution.  The larger size of the bipolar
outflows is associated with the larger $\machs$.  The size of the
bipolar outflows is correlated with the time in
$\tp\gtrsim2\times10^{3}\yr$.  As a result of fitting the data to a
power-law function using the least square method, the scaling relation
suggests that the size of the bipolar outflows is approximately
proportional to $\tp^{11/20}$.

\begin{figure}
  \begin{center}
    \includegraphics[clip,width=0.95\linewidth]{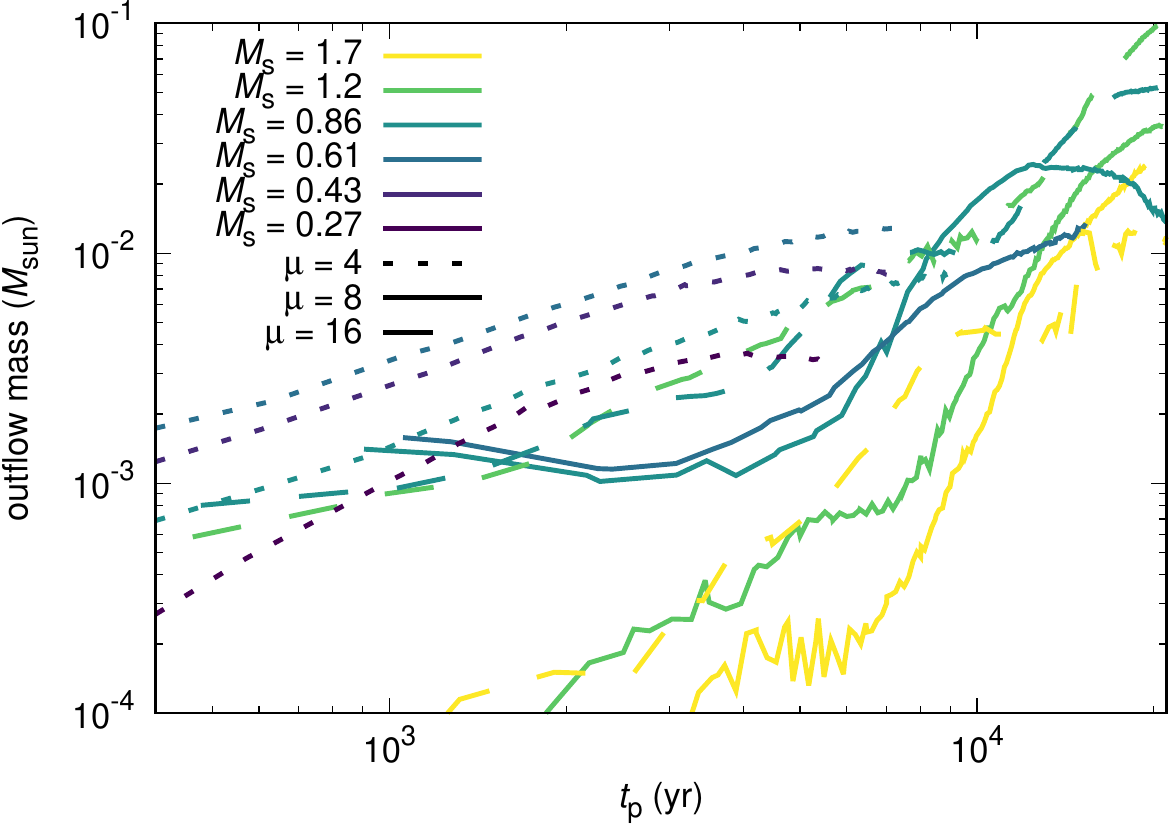}
    \includegraphics[clip,width=0.95\linewidth]{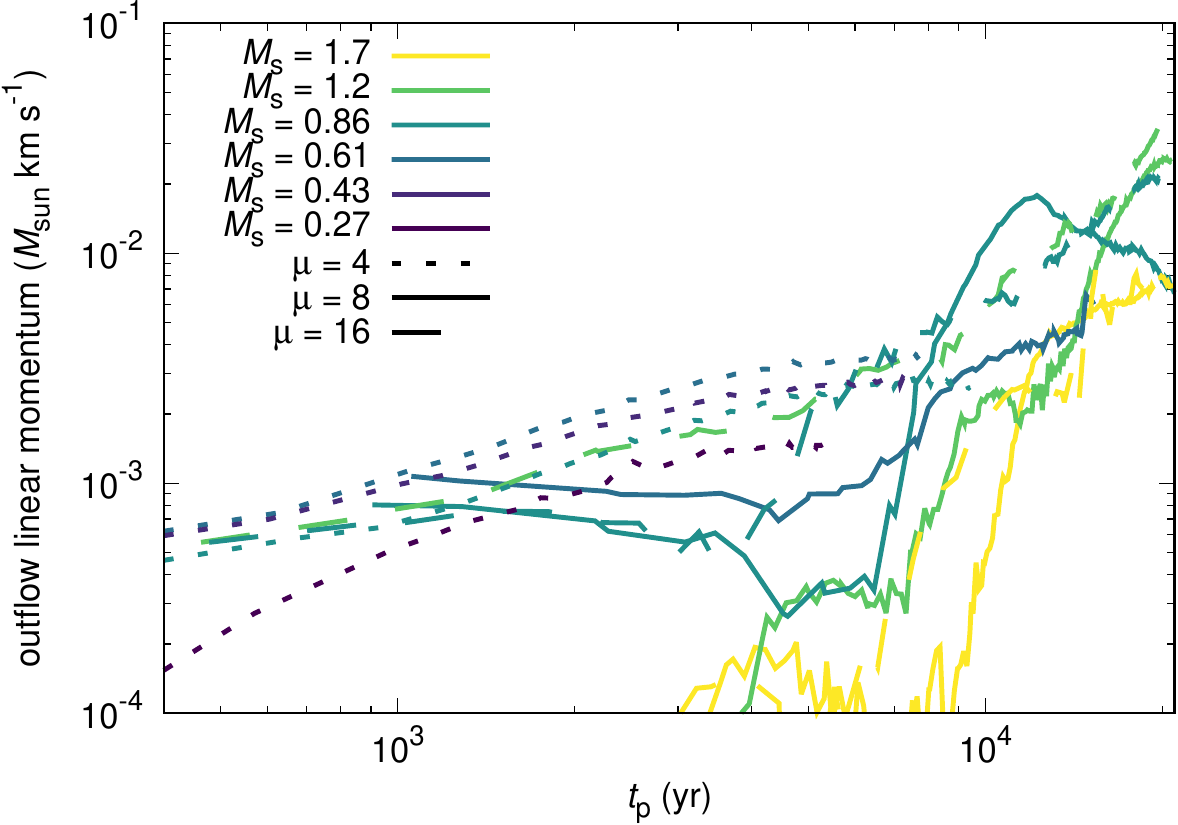}
    \includegraphics[clip,width=0.95\linewidth]{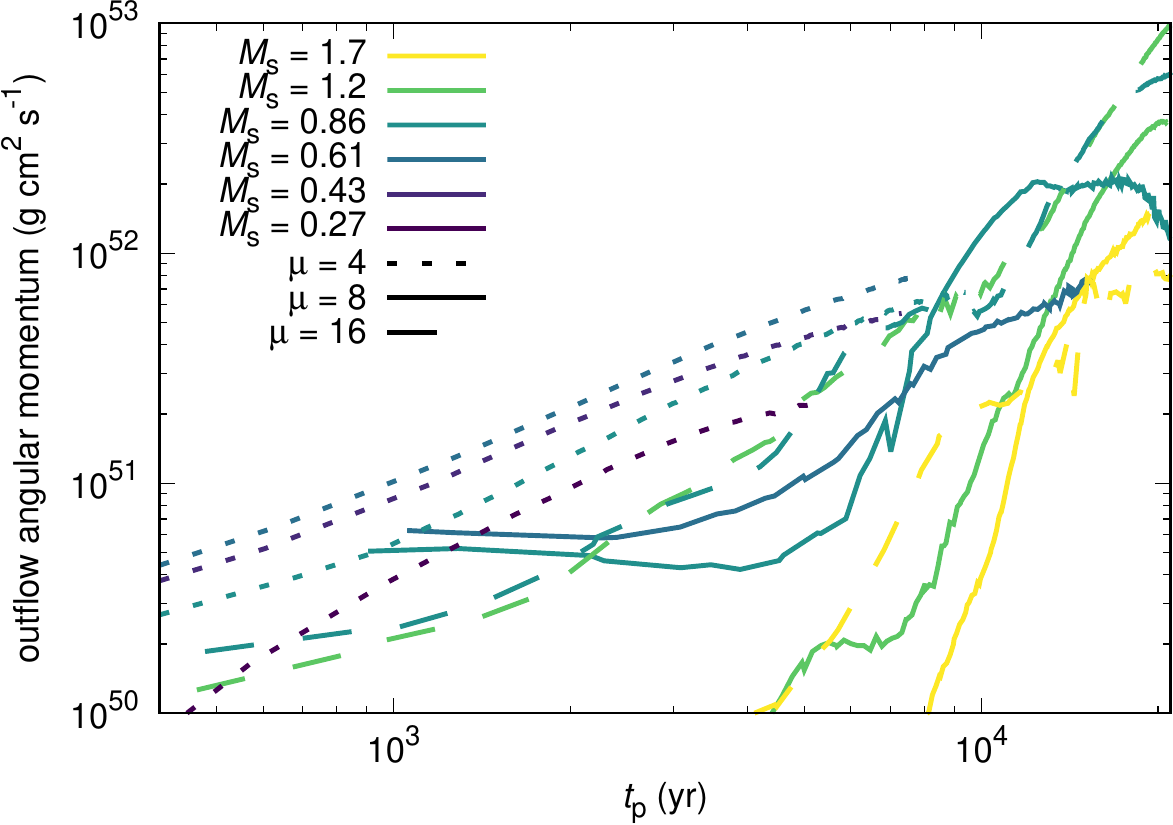}
  \end{center}
  \caption{ Time evolution of the mass $M_{\rm{out}}$ (top), linear
    momentum $|\bm{P}_{\rm{out}}|$ (middle), and angular momentum
    $|\bm{J}_{\rm{out}}|$ (bottom) for the unipolar (solid and dashed)
    and bipolar (dotted) outflows.  Line types and colors are the same
    as in Figure \ref{fig:time_evol_outflow_size}.  \\
    {Alt text: Line graphs showing the evolution of mass, linear
      momentum, and angular momentum of unipolar and bipolar
      outflows.}
  }
  \label{fig:time_evol_outflow_mass_mom_jang}
\end{figure}

Figure \ref{fig:time_evol_outflow_mass_mom_jang} shows the time
evolution of the mass $M_{\rm{out}}$, linear momentum
$|\bm{P}_{\rm{out}}|$, and angular momentum $|\bm{J}_{\rm{out}}|$ for
the unipolar and bipolar outflows, which are calculated as
\begin{align}
  M_{\rm{out}} &= {\ssum_{j=1}^{N_{\rm{out}}}} m_{{\rm{out}},j},
  \label{eq:mass_out} \\
  \bm{P}_{\rm{out}} &= {\ssum_{j=1}^{N_{\rm{out}}}} m_{{\rm{out}},j}
  (\vel_{{\rm{out}},j}-\vel_{\rm{p}})\thinspace{\rm{sgn}}((\rad_{{\rm{out}},j}-\rad_{\rm{p}})\cdot\bm{e}_{\rm{ref}}),
  \label{eq:mom_out} \\
  \bm{J}_{\rm{out}} &= {\ssum_{j=1}^{N_{\rm{out}}}} m_{{\rm{out}},j}
  (\rad_{{\rm{out}},j}-\rad_{\rm{p}}) \times (\vel_{{\rm{out}},j}-\vel_{\rm{p}}),
  \label{eq:jang_out}
\end{align}
where $N_{\rm{out}}$ is the number of the outflow particles.
$m_{{\rm{out}},j}$, $\rad_{{\rm{out}},j}$, and $\vel_{{\rm{out}},j}$
are the mass, position, and velocity of the $j$-th outflow particle.
The reference vector $\bm{e}_{\rm{ref}}$ is chosen here as
$\bm{e}_{\rm{ref}}=\bm{e}_{z}$, where $\bm{e}_{z}$ is the unit vector
in the direction of the $z$-axis.

As shown in Figure \ref{fig:time_evol_outflow_mass_mom_jang},
$M_{\rm{out}}$, $|\bm{P}_{\rm{out}}|$, and $|\bm{J}_{\rm{out}}|$ of
the unipolar outflows increase as time proceeds.  $M_{\rm{out}}$
rapidly increases after $\tp\sim10^{4}\yr$, and eventually reaches
$\gtrsim10^{-2}\msun$.  The mean velocity of the unipolar outflows,
defined as $|\bm{P}_{\rm{out}}|/M_{\rm{out}}$, grows to approximately
a few $\rm{km}\thinspace\rm{s}^{-1}$ in the plotted unipolar models.
The unipolar outflows eventually grow to have
$|\bm{J}_{\rm{out}}|\gtrsim10^{52}\gcmcms$.  There is a weak tendency
for $M_{\rm{out}}$, $|\bm{P}_{\rm{out}}|$, and $|\bm{J}_{\rm{out}}|$
of the unipolar outflows to be smaller as $\machs$ is larger at the
same $\tp$.

In the bipolar outflows, $M_{\rm{out}}$, $|\bm{P}_{\rm{out}}|$, and
$|\bm{J}_{\rm{out}}|$ increase gradually throughout the simulations.
The growth behavior of these quantities in the bipolar outflows is
consistent with that reported in previous simulation studies (e.g.,
\cite{2013MNRAS.431.1719M, 2020ApJ...896..158T, 2023MNRAS.521.2661K}).
The outflow mass $M_{\rm{out}}$ is in the range $10^{-3}\msun \lesssim
M_{\rm{out}} \lesssim 10^{-2}\msun$ for $10^{3}\yr \lesssim \tp
\lesssim 10^{4}\yr$, which is consistent with the observed values for
a dynamical time of $t_{\rm{dyn}}\lesssim 10^{4}\yr$
\citep{2004A&A...426..503W}.

\begin{figure}
  \begin{center}
    \includegraphics[clip,width=0.95\linewidth]{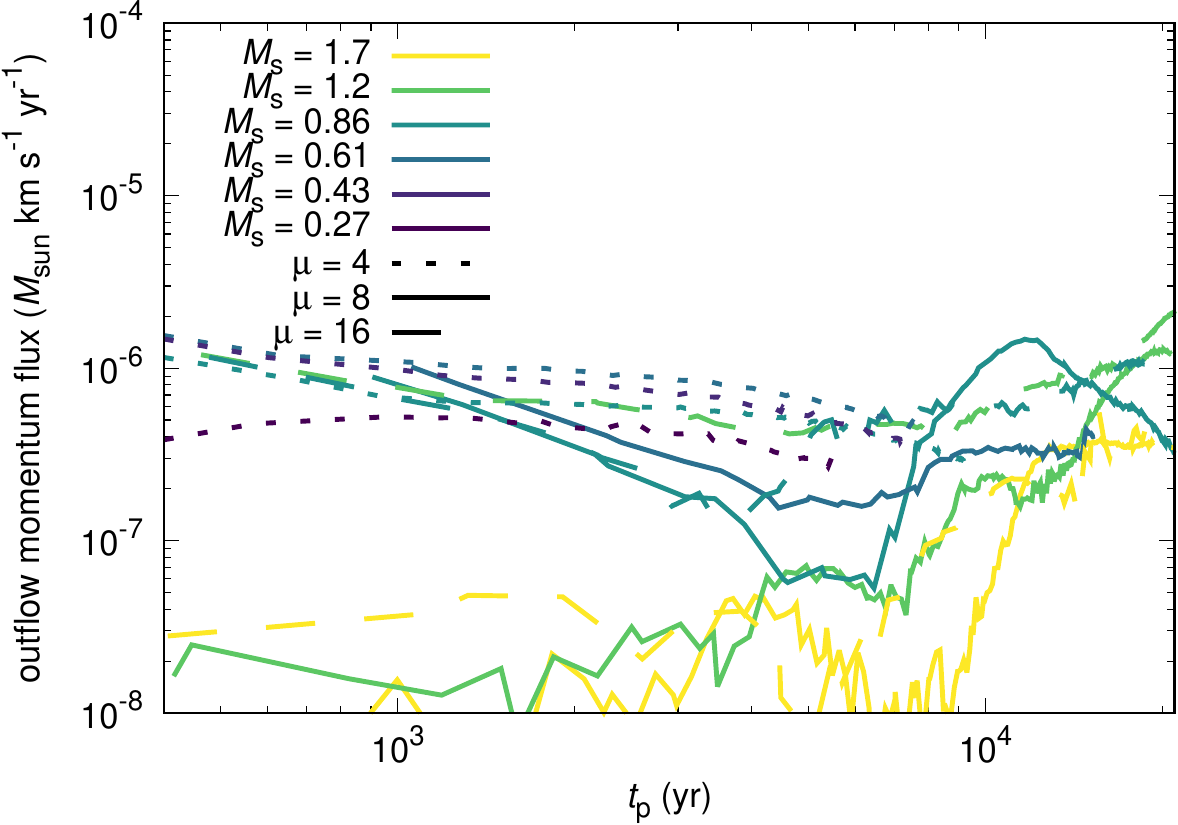}
    \includegraphics[clip,width=0.95\linewidth]{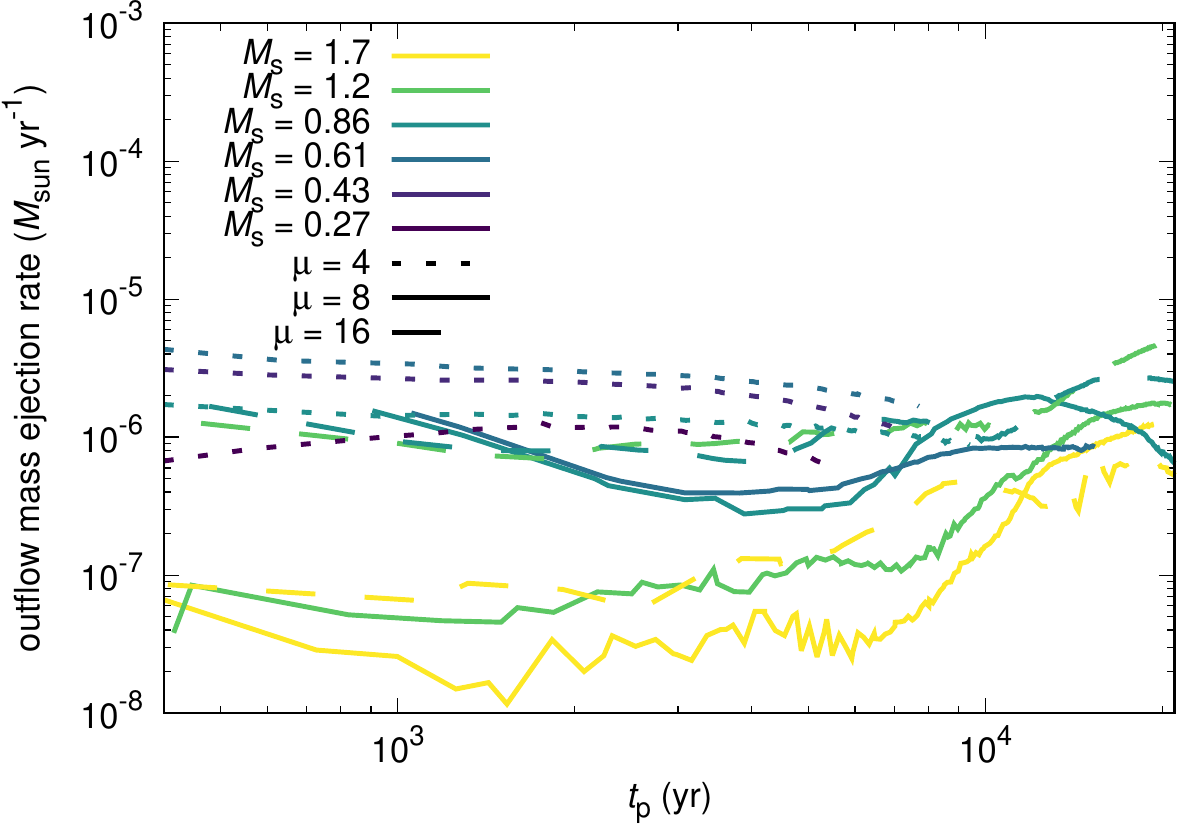}
  \end{center}
  \caption{ Time evolution of the momentum flux
    $|\bm{P}_{\rm{out}}|/\tp$ (upper) and mass ejection rate
    $M_{\rm{out}}/\tp$ (lower) for the unipolar (solid and dashed) and
    bipolar (dotted) outflows.  Line types and colors are the same as
    in Figure \ref{fig:time_evol_outflow_size}.  \\
    {Alt text: Line graphs showing the evolution of momentum flux and
      mass ejection rate of unipolar and bipolar outflows.}
  }
  \label{fig:time_evol_outflow_momFlux_massEjec}
\end{figure}

Figure \ref{fig:time_evol_outflow_momFlux_massEjec} shows the time
evolution of the momentum flux (upper panel), calculated as
$|\bm{P}_{\rm{out}}|/\tp$, and mass ejection rate (lower panel),
calculated as $M_{\rm{out}}/\tp$, for the unipolar and bipolar
outflows.  The momentum flux of the unipolar outflows exceeds
$\sim10^{-7}\msun\kms\yr^{-1}$ after $\tp\sim10^{4}\yr$.  Similarly,
the mass ejection rate becomes larger than $\sim10^{-7}\msun\yr^{-1}$
after $\tp\sim10^{4}\yr$.  The momentum flux and mass ejection rate of
the unipolar outflows tend to increase more slowly with a larger
$\machs$.  Both the momentum flux and mass ejection rate of the
bipolar outflows are $\sim10^{-6}\msun\kms\yr^{-1}$ and
$\sim10^{-6}\msun\yr^{-1}$ throughout the simulations, respectively.
This evolutionary trend of the physical quantities in the bipolar
outflows is also consistent with previous simulation results (e.g.,
\cite{2013MNRAS.431.1719M, 2020ApJ...896..158T, 2023MNRAS.521.2661K}).

\section{Discussion}
\label{sec:discussion}

\subsection{Formation of unipolar and bipolar outflows}
\label{subsec:formation_outflows}

\begin{figure*}
  \begin{center}
    \includegraphics[clip,width=\linewidth]{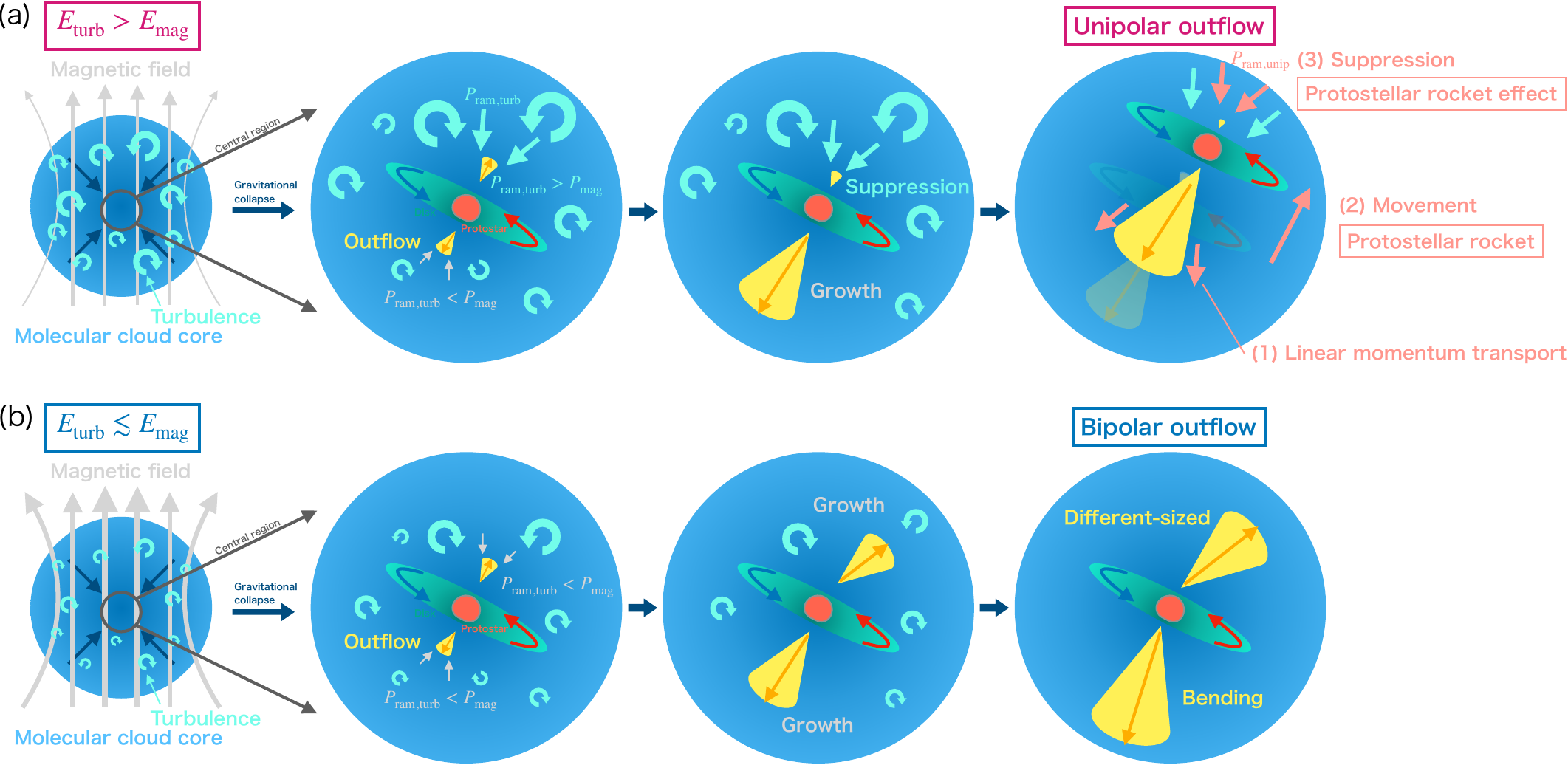}
  \end{center}
  \caption{ Schematic illustration of the formation of (a) unipolar
    and (b) bipolar outflows.  \\
    {Alt text: Schematic formation pathways.}
  }
  \label{fig:schematic_view_outflows}
\end{figure*}

In our simulations, both unipolar and bipolar outflows are formed
during the gravitational collapse of magnetized turbulent cloud cores.
Figure \ref{fig:schematic_view_outflows} illustrates the formation
scenarios for these two types of outflows.

As shown in Figure \ref{fig:schematic_view_outflows}(a), the formation
of the unipolar outflows occurs when the turbulent energy is larger
than the magnetic energy in the initial cloud cores
($E_{\rm{turb}}>E_{\rm{mag}}$).  In this case, the turbulence
dominates over the magnetic field in envelope accretion.  It causes
the accretion to become strongly asymmetric.  The gas motions
generated by this turbulent accretion regulate the magnetic field
configuration.  Consequently, the growth of one side of the initial
bipolar outflow is suppressed by the turbulent ram pressure
$P_{\rm{ram,turb}}$ that overcomes the magnetic pressure
$P_{\rm{mag}}$, as reported in \citet{2024ApJ...963...20T}.  This
suppression results in the formation of the unipolar outflow.

In the subsequent evolution, the suppression is enhanced due to the
\textit{protostellar rocket effect} \citep{2024ApJ...963...20T}.  Once
the unipolar outflow is driven, the protostellar system is gradually
accelerated by the linear momentum transport from the unipolar
outflow.  This acceleration causes the system to move toward the outer
region of the cloud core, similar to the launch and propulsion of a
rocket.  As the velocity of the system with respect to the infalling
envelope increases, the additional ram pressure $P_{\rm{ram,unip}}$ is
generated.  As a result, the suppression on the growth of the one side
is further enhanced because $P_{\rm{mag}}$ cannot overcome the
combined pressure of $P_{\rm{ram,turb}}$ and $P_{\rm{ram,unip}}$.

The formation of the bipolar outflows occurs when the turbulent energy
is smaller than or comparable to the magnetic energy in the initial
cloud cores ($E_{\rm{turb}}\lesssim E_{\rm{mag}}$), as shown in Figure
\ref{fig:schematic_view_outflows}(b).  In this case, the magnetic
field is comparable to or dominates over the turbulence in envelope
accretion, leading to weakly turbulent accretion.  Consequently, the
initial bipolar outflow is suppressed weakly and perturbed by the
envelope accretion.  It results the formation of bending and
different-sized outflows.

Here, we also find that no outflow is formed in the cloud cores with
the strong magnetic field of $\magmu=2$ ($\magmu_{\rm{const}}=1$).
The strong magnetic field rapidly extracts the angular momentum from
the central region of the cloud core due to the magnetic braking,
resulting in the suppression of Keplerian disk formation.
\citet{2018MNRAS.473.3080M} investigated star formation in
magnetically subcritical rotating cloud cores (corresponding to
$\magmu_{\rm{const}}\geq 1$) using three-dimensional non-ideal MHD
simulations.  They found that no large-scale low-velocity outflow
appears in the collapsing subcritical cloud cores because the angular
momentum of the core is sufficiently transferred to the interstellar
medium by the magnetic braking before the rapid gravitational collapse
begins.  Our results are consistent with their findings.

\subsection{Formation condition of unipolar outflows}
\label{subsec:condi_unip}

\begin{figure}
  \begin{center}
    \includegraphics[clip,width=0.95\linewidth]{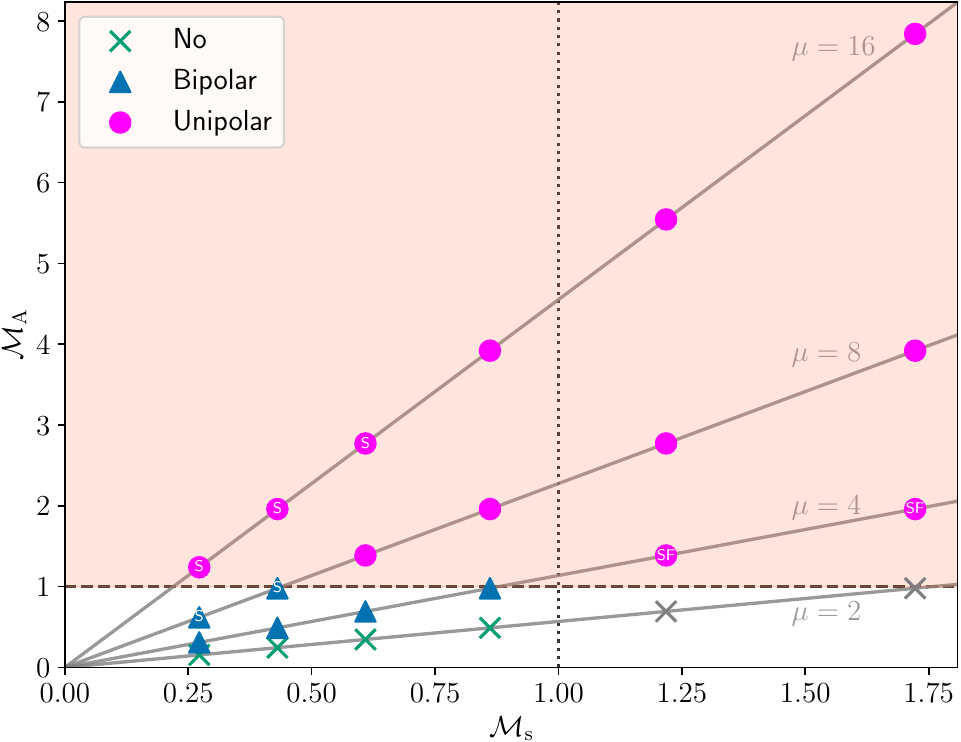}
  \end{center}
  \caption{ Outflow morphologies plotted on
    $(\machs,\thinspace\machA)$ plane.  The circles (magenta),
    triangles (blue), and crosses (green and gray) indicate the
    unipolar, bipolar, and no outflows.  Shaded region shows
    $\machA>1$.  Capitals ``S'' and ``F'' mean the small
    ($\lesssim5\times10^{2}\au$) and failed outflows.  \\
    {Alt text: Scatter plot of simulation results on
      $(\machs,\thinspace\machA)$ plane.}
  }
  \label{fig:outflow_morphologies_plane_Mach}
\end{figure}

Our results suggest that the strengths of the turbulence and magnetic
field of the cloud cores play a key role in the formation of unipolar
outflows.  The relative strength of the turbulence to the magnetic
field can be expressed by the Alfv\'{e}n Mach number of the initial
cloud core, defined as
\begin{align}
  \machA
  &\equiv \sqrt{\frac{E_{\rm{turb}}}{E_{\rm{mag}}}}
  = \frac{\sqrt{3}\sigma_{\rm{v}}}{\overline{v}_{\rm{A}}}
  = \left( \frac{\alpha}{3\gmag} \right)^{\frac{1}{2}} \machs \nonumber \\
  &=
  \left( \frac{\pi^2 G}{12} \frac{\Lambda_{\rm{grav}}}{\Lambda_{\rm{mag}}} \right)^{\frac{1}{2}}
  \left( \frac{M_{\rm{c}}}{\Phi_{\rm{mag}}} \right)_{\rm{crit}}
  \alpha^{\frac{1}{2}} \magmu \machs,
  \label{eq:alf_mach}
\end{align}
where $\overline{v}_{\rm{A}}$ is the root-mean-square Alfv\'{e}n
velocity of the initial cloud core.  The constants of
$\Lambda_{\rm{grav}}$ and $\Lambda_{\rm{mag}}$ are calculated by
equations (\ref{app:eq:Lgrav}) and (\ref{app:eq:Lmag}), respectively
(see the appendix \ref{app:sec:pq_core} for more details).

Figure \ref{fig:outflow_morphologies_plane_Mach} plots the outflow
morphologies on the $\machs-\machA$ plane.  This figure clearly
indicates that the unipolar outflows are formed in the initial cloud
cores under conditions of the super-Alfv\'{e}nic turbulence:
\begin{eqnarray}
  \machA > 1.
  \label{eq:machA_cond_unip}
\end{eqnarray}

Substituing equation (\ref{eq:alf_mach}) into inequality
(\ref{eq:machA_cond_unip}), the formation condition of the unipolar
outflows can be written as
\begin{eqnarray}
  \left(\frac{\alpha}{0.5}\right)^{\frac{1}{2}}
  \left(\frac{\magmu}{5}\right)
  \left(\frac{{\cal{M}_{\rm{s}}}}{1.0}\right)
  >
  0.63.
  \label{eq:machA_cond_unip_mod}
\end{eqnarray}

Inequality (\ref{eq:machA_cond_unip_mod}) enables us to predict the
formation of unipolar outflows based on the observed physical
quantities of pre-stellar cores, $\alpha$, $\machs$, and $\magmu$.  In
other words, the observed outflow morphologies can be used as tracers
to estimate the physical properties of the parent cloud cores.
Furthermore, the statistical analysis of the observed outflow
morphologies could clarify the relative strength of the turbulence to
the magnetic field in the star formation environments.  This will
provide new insights into whether the unipolar outflows are the
apparent effects of observation or intrinsic phenomena in the star
formation process.

Here, we note that condition (\ref{eq:machA_cond_unip_mod}) depends on
$\alpha$, which represents the strength of gravity of cloud cores and
is linked to the formation of the binary and multiple protostellar
systems.  The self-gravitating gas tends to fragment at lower values
of $\alpha$, both before and after the protostar formation (e.g.,
\cite{1979ApJ...234..289B, 1980ApJ...235..866T, 1980ApJ...242..209B,
  1981ApJ...248..717T, 1984ApJ...279..621M, 1992PASJ...44..193M,
  1993ApJ...410..157B, 1999ApJ...523L.155T, 1999ApJ...526..307T}),
although the fragmentation during the cloud core collapse depends on
the ratio of the rotational to gravitational energies $\brot$ (e.g.,
\cite{2003MNRAS.340...91C, 2003ApJ...595..913M, 2011MNRAS.416..591T,
  2015MNRAS.446.1175T}) and the magnetic field (e.g.,
\cite{2004MNRAS.348L...1M, 2007MNRAS.377...77P, 2008ApJ...677..327M,
  2017MNRAS.466.1788W, 2019MNRAS.486.2587W}).  However, the rigorous
conditions for the fragmentation remain unclear in magnetized
turbulent cloud cores to date.  We plan to study the fragmentation
condition in magnetized turbulent cloud cores and whether the
formation condition represented by condition
(\ref{eq:machA_cond_unip_mod}) can be applied to the binary and
multiple systems.

The present study assumes that the net angular momentum $\Jcore$ and
magnetic field of the initial cloud core are initially parallel.
Previous studies have shown that the misalignment between the initial
angular momentum and magnetic field can influence disk formation and
outflow activity even in rigidly rotating cloud cores (e.g.,
\cite{2012A&A...543A.128J, 2013ApJ...774...82L, 2018ApJ...868...22T}).
Nevertheless, we intentionally focus on the aligned configuration in
order to clarify the role of the balance between turbulent and
magnetic energies.  Since such a configuration is generally considered
favorable for bipolar outflow formation, the present result
demonstrates that sufficiently strong turbulence can nevertheless
drive unipolar outflows.  A systematic investigation of the dependence
on the misalignment will be presented in future work.

\subsection{Polarity of outflows}
\label{subsec:polarity}

The outflows have a wide range of physical properties.  The outflow
morphologies are different in each simulation, such as the unipolar
and bipolar outflows.  To identify and quantify the diversity of the
outflows, we utilize an analysis using the spherical harmonics
expansion, which is usually applied to characterize anisotropies in
the cosmic microwave background (CMB) radiation (see
\cite{2002ApJ...567....2H} for instance).

We apply the spherical harmonics expansion to the outflow as
\begin{align}
  \Delta \hat{P}_{\rm{out}}(\nhat)
  &\equiv \frac{P_{\rm{out}}(\theta,\phi)-\bar{P}_{\rm{out}}}{\bar{P}_{\rm{out}}}
  = {\ssum_{l=1}^{\infty}} {\ssum_{m=-l}^{m=l}} a_{lm} Y_{l}^{m}(\theta,\phi),
  \label{eq:delta_mom_out} \\
  P_{\rm{out}}(\theta,\phi)
  &\equiv |\bm{P}_{\rm{out}}(\theta,\phi)|,
  \label{eq:mom_out_abs} \\
  \bar{P}_{\rm{out}}
  &\equiv \int_{4\pi} \frac{d\Omega}{4\pi} P_{\rm{out}}(\theta,\phi) \nonumber \\
  &= \int_{-1}^{1} \frac{d(\cos\theta)}{2}
  \int_{0}^{2\pi} \frac{d\phi}{2\pi}
  |\bm{P}_{\rm{out}}(\theta,\phi)|,
  \label{eq:mom_out_mean}
\end{align}
where $a_{lm}$ are the expansion coefficients,
$Y_{l}^{m}(\theta,\phi)$ are the spherical harmonics,
$|\bm{P}_{\rm{out}}(\theta,\phi)|$ is the linear momentum distribution
of the outflow, and $d\Omega=\sin\theta d\theta d\phi$ is the
differential solid angle in spherical coordinates (see the appendix
\ref{app:spherical_harmonics} for more details).

\begin{figure*}
  \begin{center}
    \includegraphics[clip,width=0.80\linewidth]{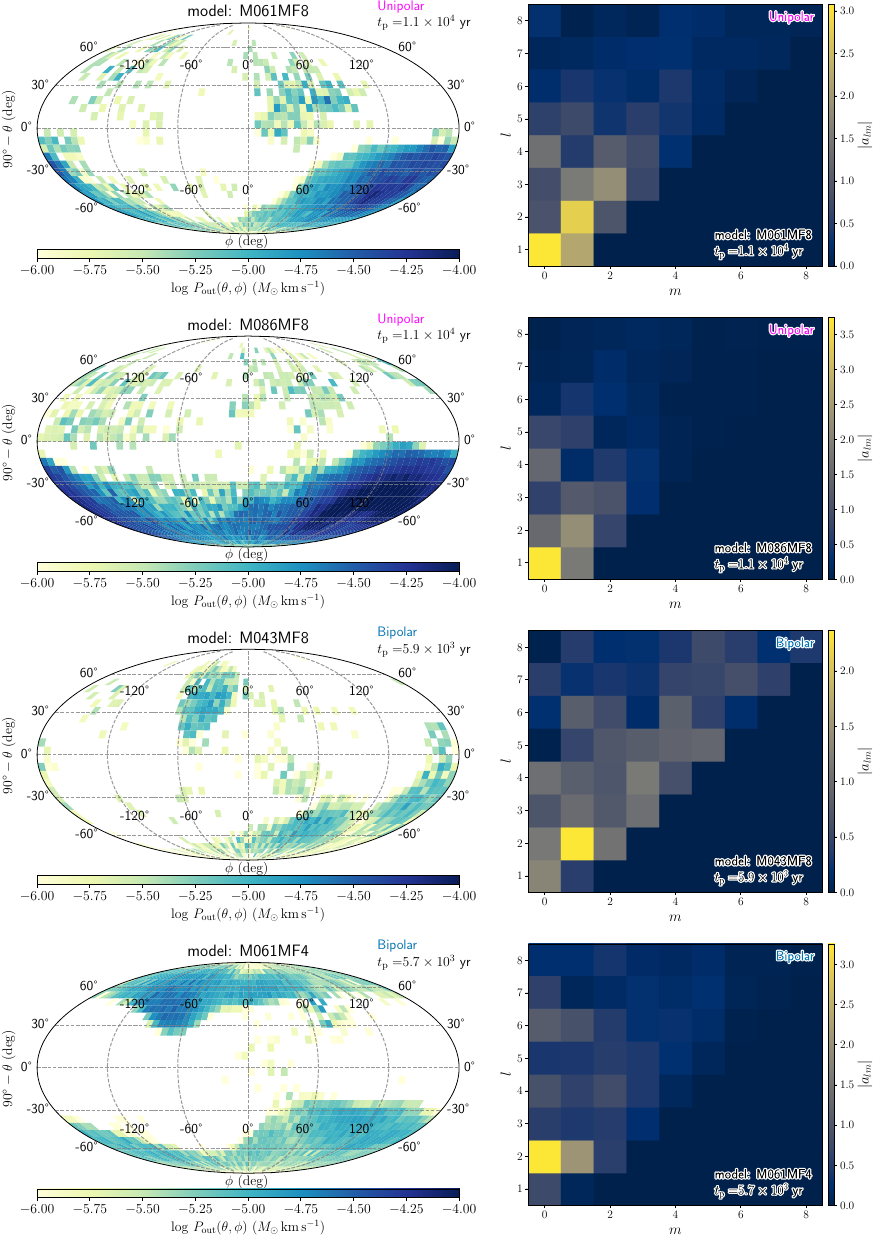}
  \end{center}
  \caption{ The linear momentum distribution
    $P_{\rm{out}}(\theta,\phi)$ (left panels) and the expansion
    coefficients $|a_{lm}|$ (right panels) of the unipolar and bipolar
    outflows for the models M061MF8 ($\machs=0.61$, $\magmu=8$),
    M086MF8 ($\machs=0.86$, $\magmu=8$), M043MF8 ($\machs=0.43$,
    $\magmu=8$), and M061MF4 ($\machs=0.61$, $\magmu=4$), which are
    the same as in Figure \ref{fig:surface_density_map}.  The left
    panels are calculated with the cell numbers of $N_{\theta}=32$ and
    $N_{\phi}=64$, corresponding to the cell widths of
    $\Delta\theta=\pi/N_{\theta}$ and $\Delta\phi=2\pi/N_{\phi}$.  The
    right panels are calculated in $0\leq l\leq l_{\rm{max}}$, with
    $l_{\rm{max}}=75$ and the cell numbers of $2N_{\theta}=64$ and
    $2N_{\phi}=128$.  \\
    {Alt text: Maps of linear momentum distribution and spherical
      harmonics expansion coefficients.}
  }
  \label{fig:polarity_mom_alm}
\end{figure*}

Figure \ref{fig:polarity_mom_alm} shows the linear momentum
distribution, $P_{\rm{out}}(\theta,\phi)$ (left panels), and the
expansion coefficients, $|a_{lm}|$ (right panels), for the same models
shown in Figure \ref{fig:surface_density_map}.  In the right panels,
the unipolar outflows exhibit a peak in $|a_{lm}|$ at $l=1$ (dipole
term), whereas the bipolar outflows show a maximum $|a_{lm}|$ at $l=2$
(quadrupole term).  This indicates that the outflow morphologies can
be quantitatively classified by the value of $l$ for the maximum of
$|a_{lm}|$.  We have confirmed that the outflow morphologies visually
classified in Figure \ref{fig:3Dview} are consistent with this
classification scheme.  The degree of the inclination angle for the
outflow can be characterized by the value of $m$.

\begin{figure}
  \begin{center}
    \includegraphics[clip,width=0.95\linewidth]{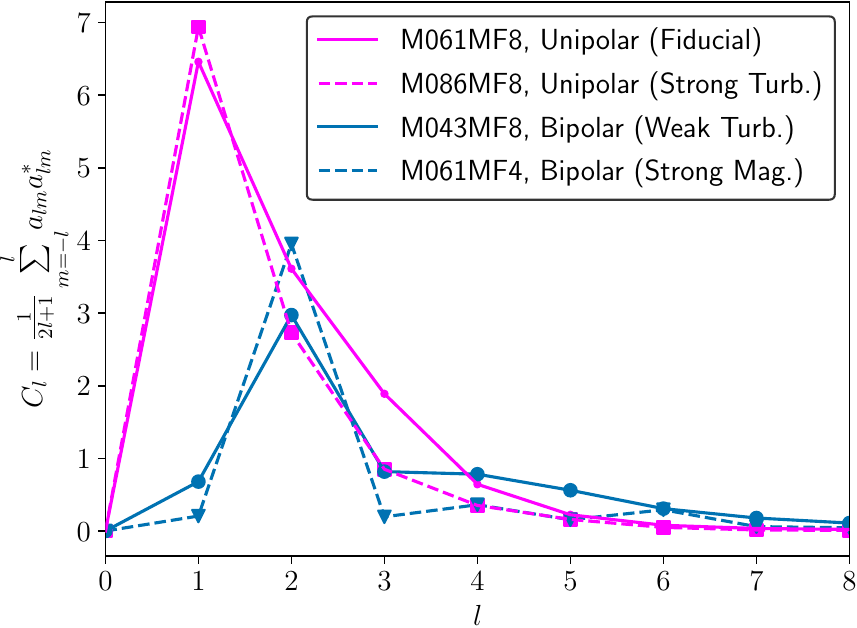}
  \end{center}
  \caption{ The polarity of outflows $C_{l}$ for unipolar and bipolar
    outflows, for the same models shown in Figure
    \ref{fig:polarity_mom_alm}.  $C_{l}$ are calculated in $0\leq
    l\leq l_{\rm{max}}$, with $l_{\rm{max}}=75$ and the cell numbers
    of $2N_{\theta}=64$ and $2N_{\phi}=128$.  \\
    {Alt text: Line graphs showing the polarity of outflows.}
  }
  \label{fig:polarity_power_spec_1D}
\end{figure}

Although the outflows can be classified by their dominant multipole
order $l$, the contributions from other multipole components differ in
each simulation even in the same classified morphologies.  Here, to
characterize these differences more clearly, we introduce the concept
of the \textit{polarity of outflows}, $C_{l}$, which is expressed by
the same definition as the angular power spectrum used in the CMB
radiation analysis:
\begin{align}
  C_{l} = {\frac{1}{2l+1}\textstyle\sum\limits_{m=-l}^{l}} a_{lm} a_{lm}^{*}.
  \label{eq:C_l}
\end{align}

Figure \ref{fig:polarity_power_spec_1D} shows the polarity of outflows
for the same models shown in Figure \ref{fig:polarity_mom_alm}.
Comparing the unipolar outflows, $C_{1}$ of the model M086MF8 (dashed
magenta line) is larger than that of the model M061MF8 (solid magenta
line).  In contrast, $C_{2}$ of the model M086MF8 is smaller than that
of the model M061MF8.  Also, for the bipolar outflows, $C_{2}$ of the
model M061MF4 (dashed blue line) is larger than that of the model
M043MF8 (solid blue line), whereas $C_{1}$, $C_{3}$, and several of
$C_{l}$ of the model M061MF4 are smaller than that of the model
M043MF8.  These results clearly demonstrate that the distribution of
$C_{l}$ effectively characterizes the differences in the same
classified morphologies.  This indicates that the profile of $C_{l}$
can be used to describe the \textit{unipolarity} and
\textit{bipolarity} of outflows.

\subsection{Bending structure in bipolar outflows}
\label{subsec:asym_feature}

Recent observations of the Class 0/I YSOs have revealed the bending
structures in bipolar outflows and possible changes in outflow
directions with time (e.g., \cite{2017ApJ...834..178Y,
  2018ApJ...863...19A, 2019ApJ...887..209A, 2023ApJ...947...25H,
  2023ApJ...953..190K, 2024ApJ...966..192S, 2024AJ....168..107L,
  2025ApJ...995..225C}).  As shown in Figure \ref{fig:3Dview}, our
simulations naturally reproduce such bending outflows, primarily due
to the turbulent accretion of the envelope.  This suggests that the
observed bending structures in bipolar outflows may be closely linked
to, and explained by, the turbulent nature of the infalling envelope.

\begin{figure}
  \begin{center}
    \includegraphics[clip,width=\linewidth]{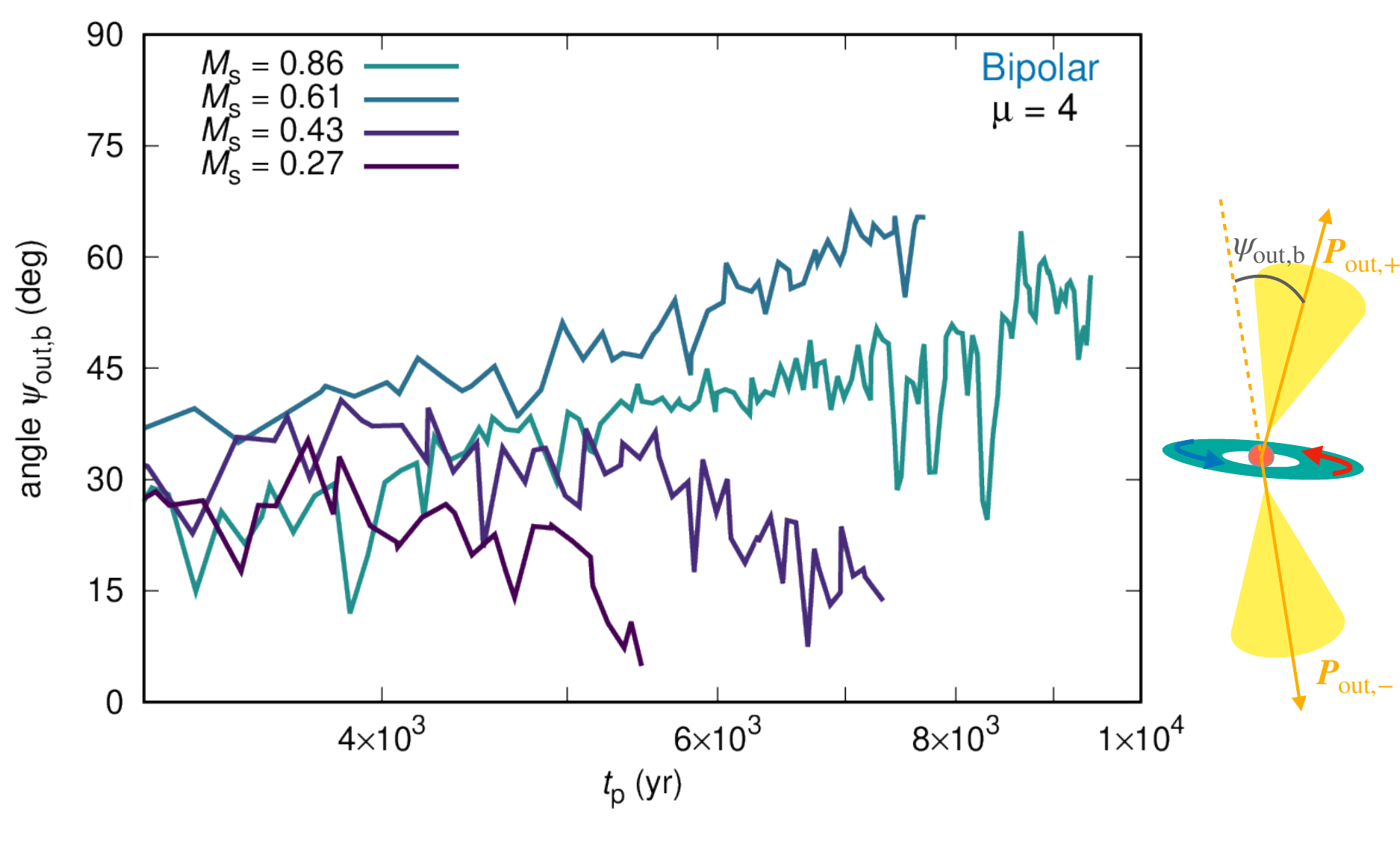}
  \end{center}
  \caption{ Time evolution of the bending angle $\psi_{\rm{out,b}}$
    for the bipolar outflows with $\magmu=4$ after
    $\tp=3\times10^{3}\yr$.  The line colors are the same as in Figure
    \ref{fig:time_evol_outflow_size}.  \\
    {Alt text: Line graphs showing the evolution of bending angle.}
  }
  \label{fig:time_evol_bending_angle}
\end{figure}

In Figure \ref{fig:time_evol_bending_angle}, we plot the time
evolution of the bending angle $\psi_{\rm{out,b}}$ defined as
\begin{eqnarray}
  \psi_{\rm{out,b}} =
  \cos^{-1}\left(
  -\frac{\bm{P}_{\rm{out,+}}\cdot\bm{P}_{\rm{out,-}}}
  {|\bm{P}_{\rm{out,+}}||\bm{P}_{\rm{out,-}}|}
  \right),
  \label{eq:bending_angle}
\end{eqnarray}
where $\bm{P}_{\rm{out,+}}$ and $\bm{P}_{\rm{out,-}}$ denote the
linear momentum of the outflow in the regions $z>0$ and $z<0$,
respectively, corresponding to the two components of the bipolar
outflow.

As shown in Figure \ref{fig:time_evol_bending_angle}, the bending
angle with larger initial sonic Mach number exhibits a greater
increase as time proceeds.  In the growth cases, the value of
$\psi_{\rm{out,b}}$ reaches approximately $60^{\circ}-65^{\circ}$,
suggesting that the degree of bending may serve as an observational
indicator of the strength of the turbulence in the parent cloud core.
Our results support the possibility that the bending structures
observed in bipolar outflows may originate from the turbulent
dynamics.

\section{Summary}
\label{sec:summary}

This study investigates the impact of the turbulence and magnetic
field on the formation of unipolar outflows during the collapse of
isolated low-mass cloud cores, covering a wide range of initial
parameters, $\machs$ and $\magmu$.  We summarize our results and
findings as follows.

\begin{enumerate}
\item[1.]  The unipolar outflows are formed in the initial cloud cores
  with $E_{\rm{turb}}>E_{\rm{mag}}$, i.e., with the super-Alfv\'{e}nic
  turbulence of ${\cal{M}_{\rm{A}}}>1$, in the conditions explored in
  this study.
\item[2.] The bipolar outflows are formed in the initial cloud cores
  with the sub-Alfv\'{e}nic and trans-Alfv\'{e}nic turbulence of
  ${\cal{M}_{\rm{A}}}\lesssim 1$.
\item[3.]  Outflows are not formed in the initial cloud cores with the
  strong magnetic field of $\magmu=2$ ($\magmu_{\rm{const}}\sim1$).
\item[4.]  The strengths of the turbulence and magnetic field of the
  cloud cores play a crucial role in the formation of unipolar
  outflows.  The results suggest that the outflow morphologies can be
  used as tracers to estimate the strengths of the turbulence and
  magnetic field of parent cloud cores indirectly.
\item[5.]  The outflows can be characterized by the analysis using the
  spherical harmonics expansion.  We find that the \textit{polarity of
    outflows} are well described by the distribution of $C_{l}$.
\item[6.]  Bending bipolar outflows are naturally formed.  The bending
  angles roughly depend on the strength of the turbulence, suggesting
  that the observed bending structures may be originated from the
  turbulent dynamics.
\end{enumerate}



\begin{ack}
  We thank Dr. Tomoaki Matsumoto, Dr. Masahiro Machida, and
  Dr. Keiichi Wada for the comments and fruitful discussions.
  We are also grateful to the anonymous referee for
  the fruitful suggestions and insightful comments that helped to
  improve the manuscript.
  Numerical computations were carried out on Cray XC50 and XD2000 at
  Center for Computational Astrophysics, National Astronomical
  Observatory of Japan.  D.T. acknowledges the fellowship by the Japan
  Society for the Promotion of Science (JSPS).
  This work is supported by JSPS KAKENHI Grant Numbers
  JP21J23102 (D.T.),
  JP24K22906 (D.T.),
  JP21H00048 (S.T.),
  JP21H04495 (S.T.),
  JP24K00674 (S.T.),
  JP18H01247 (Y.S.) and
  JP23K25908 (Y.S.).
  Y.T. is supported by JST FOREST Grant Number JPMJFR2234.
  S.T. is supported by NAOJ ALMA Scientific Research Grant Number 2022-20A.
\end{ack}


\section*{Data availability}
The data underlying this article will be shared on reasonable request
to the corresponding author.

\appendix 
\section{Magnetic and gravitational energies of the initial cloud core:
  scaling relation between $\gmag$ and $\magmu$}
\label{app:sec:pq_core}

The non-dimensional density profile of the Bonnor-Ebert sphere
$\varrho_{\rm{BE}}$ is obtained by solving the equation for an
isothermal gas sphere in gravitational equilibrium
\citep{1939isss.book.....C}:
\begin{equation}
  \frac{1}{\xi^2}\frac{d}{d\xi}
  \left(\xi^2\frac{d\psi}{d\xi}\right)
  = \exp[-\psi],
  \label{app:eq:hydrostatic_eq_iso_selfgrav}
\end{equation}
where $\psi$ and $\xi$ are defined by
\begin{align}
  \psi &= - \ln \varrho_{\rm{BE}}, \label{app:eq:variable1} \\
  \xi  &= r\sqrt{\frac{4\pi G \rho_{\rm{c}}}{c_{\rm{s,iso}}^2}}
  = \frac{r}{a}. \label{app:eq:variable2}
\end{align}
The boundary conditions are imposed as follows:
\begin{align}
  \left. \psi \right|_{\xi=0} &= 0, \\
  \left. \frac{d\psi}{d\xi} \right|_{\xi=0} &= 0.
  \label{app:eq:boundary}
\end{align}
The critical Bonnor-Ebert sphere has a maximum radius of
$\xi_{\rm{max}}=6.45$, meaning that
$r_{\rm{max}}=\xi_{\rm{max}}a=R_{\rm{c}}$.

The enclosed mass of the initial cloud core is calculated as
\begin{align}
  M(r)
  &= \int_{0}^{r} 4\pi \rho(r) r^2 dr \nonumber \\
  &= 4\pi f\rho_{\rm{c}} a^3 \int_{0}^{\xi} \varrho_{\rm{BE}}(\xi) \xi^2 d\xi \nonumber \\
  &= 4\pi f\rho_{\rm{c}} a^3 \int_{0}^{\xi} \exp[-\psi(\xi)] \xi^2 d\xi \nonumber \\
  &= 4\pi f\rho_{\rm{c}} a^3 \xi^2  \left.\frac{d\psi}{d\xi}\right|_{\xi}.
  \label{app:eq:Menclose}
\end{align}

Using the above equations, the gravitational energy of the initial
cloud core $E_{\rm{grav}}$ is obtained as follow:
\begin{align}
  E_{\rm{grav}}
  &= - \int_{0}^{R_{\rm{c}}} \frac{GM(r)}{r} dM(r) \nonumber \\
  &= - \frac{G (4\pi f\rho_{\rm{c}} a^3)^2}{a}
  \int_{0}^{\xi_{\rm{max}}}
  \frac{\xi^2\frac{d\psi}{d\xi}}{\xi}
  \exp[-\psi(\xi)] \xi^2 d\xi \nonumber \\
  &= - \frac{G M_{\rm{c}}^2}{R_{\rm{c}}}
  \frac{\xi_{\rm{max}}}{\left( \xi_{\rm{max}}^2 \left. \frac{d\psi}{d\xi} \right|_{\xi_{\rm{max}}} \right)^2}
  \int_{0}^{\xi_{\rm{max}}}
  \frac{d\psi}{d\xi} \exp[-\psi] \xi^3 d\xi \nonumber \\
  &= - \Lambda_{\rm{grav}} \frac{G M_{\rm{c}}^2}{R_{\rm{c}}}, \label{app:eq:Egrav}
\end{align}
where $\Lambda_{\rm{grav}}$ is a constant obtained by integrating as
\begin{align}
  \Lambda_{\rm{grav}}
  &= \frac{1}{\xi_{\rm{max}}^3 \left.\frac{d\psi}{d\xi}\right|_{\xi_{\rm{max}}}^2}
  \int_{0}^{\xi_{\rm{max}}}
  \frac{d\psi}{d\xi} \exp[-\psi] \xi^3 d\xi \nonumber \\
  &= \frac{1}{\xi_{\rm{max}}^3 \left.\frac{d\psi}{d\xi}\right|_{\xi_{\rm{max}}}^2}
  \int_{0}^{\xi_{\rm{max}}}
  \frac{d}{d\xi}\left( - \exp[-\psi] \right) \xi^3 d\xi \nonumber \\
  &= \frac{1}{\xi_{\rm{max}}^3 \left.\frac{d\psi}{d\xi}\right|_{\xi_{\rm{max}}}^2} \nonumber \\
  &\ \ \ \  \times \left\{
  \left[-\exp[-\psi]\xi^3\right]_{0}^{\xi_{\rm{max}}}
  - \int_{0}^{\xi_{\rm{max}}} ( -3\exp[-\psi]\xi^2 ) d\xi
  \right\} \nonumber \\
  &= \frac{1}{\xi_{\rm{max}}^3 \left.\frac{d\psi}{d\xi}\right|_{\xi_{\rm{max}}}^2} \nonumber \\
  &\ \ \ \  \times \left\{
  \left[-\exp[-\psi]\xi^3\right]_{0}^{\xi_{\rm{max}}}
  +3\left[\xi^2 \frac{d\psi}{d\xi}\right]_{0}^{\xi_{\rm{max}}}
  \right\} \nonumber \\
  &= \left.\frac{d\psi}{d\xi}\right|_{\xi_{\rm{max}}}^{-2}
  \left(
  -\exp[-\psi(\xi_{\rm{max}})]
  +3\xi_{\rm{max}}^{-1} \left.\frac{d\psi}{d\xi}\right|_{\xi_{\rm{max}}}
  \right).
  \label{app:eq:Lgrav}
\end{align}
Here we note that the ratio of the thermal to gravitational energies
$\alpha$ can be calculated with $\Lambda_{\rm{grav}}$ as
\begin{align}
  \alpha = \frac{E_{\rm{thm}}}{|E_{\rm{grav}}|}
  &= \frac{3 R_{\rm{c}} c_{\rm{s,iso}}^2}{2\Lambda_{\rm{grav}} G M_{\rm{c}}} \nonumber \\
  &= \frac{3}{2\Lambda_{\rm{grav}} \xi_{\rm{max}} \left.\frac{d\psi}{d\xi}\right|_{\xi_{\rm{max}}}} \frac{1}{f} \nonumber \\
  &\propto f^{-1}.
  \label{app:eq:alphaf}
\end{align}

The initial cloud core has the axisymmetric magnetic field described
as the following in cylindrical coordinates $(R,\thinspace
\varphi,\thinspace z)$ (see \cite{2020ApJ...896..158T} in detail):
\begin{align}
  B_{R}(R,z) &= \frac{B_{\rm{c}}}
  {\left[1+\left(R/R_{\rm{c}}\right)^2+\left(z/R_{\rm{c}}\right)^2\right]^2}
  \left(\frac{R}{R_{\rm{c}}}\right)
  \left(\frac{z}{R_{\rm{c}}}\right), \label{app:eq:BR} \\
  B_{z}(R,z) &= \frac{B_{\rm{c}}}
  {\left[1+\left(R/R_{\rm{c}}\right)^2+\left(z/R_{\rm{c}}\right)^2\right]^2}
  \left\{1+\left(\frac{z}{R_{\rm{c}}}\right)^2\right\}. \label{app:eq:Bz}
\end{align}
Using equations (\ref{app:eq:BR}) and (\ref{app:eq:Bz}),
the magnetic energy of the initial cloud core $E_{\rm{mag}}$ is calculated as
\begin{align}
  E_{\rm{mag}}
  &= \frac{1}{8\pi}
  \int_{\varphi=0}^{\varphi=2\pi}
  \int_{R=0}^{R=R_{\rm{c}}}
  \int_{z=-\sqrt{R_{\rm{c}}^2-R^2}}^{z=\sqrt{R_{\rm{c}}^2-R^2}}
  \left( B_{R}^2+B_{z}^2 \right)  Rd\varphi \thinspace dR \thinspace dz \nonumber  \\
  &= \Lambda_{\rm{mag}} B_{\rm{c}}^2 R_{\rm{c}}^3,
  \label{app:eq:Emag}
\end{align}
where $\Lambda_{\rm{mag}}$ is a constant obtained by integrating as
\begin{align}
  \Lambda_{\rm{mag}}
  &= \frac{1}{8\pi}
  \int_{\varphi=0}^{\varphi=2\pi}
  \int_{\hat{R}=0}^{\hat{R}=1}
  \int_{\hat{z}=-\sqrt{1-\hat{R}^2}}^{\hat{z}=\sqrt{1-\hat{R}^2}}
  \hat{R} d\varphi \thinspace d\hat{R} \thinspace d\hat{z} \nonumber  \\
  &\ \ \ \  \times
  \frac{\hat{R}^2 \hat{z}^2 + \left(1+\hat{z}^2\right)^2}{\left[1+\hat{R}^2+\hat{z}^2\right]^4} \nonumber \\
  &= \frac{1}{2}
  \int_{\hat{R}=0}^{\hat{R}=1}
  \int_{\hat{z}=0}^{\hat{z}=\sqrt{1-\hat{R}^2}}
  \frac{\hat{R}^3 \hat{z}^2 + \hat{R}\left(1+\hat{z}^2\right)^2}{\left[1+\hat{R}^2+\hat{z}^2\right]^4}
  d\hat{R} \thinspace d\hat{z},
  \label{app:eq:Lmag}
\end{align}
in which $\hat{R} \equiv R/R_{\rm{c}}$ and $\hat{z} \equiv
z/R_{\rm{c}}$ are used for the integrals.

The magnetic flux of the initial cloud core
$\Phi_{\rm{mag}}\left(R\right)$ is calculated as
\begin{equation}
  \Phi_{\rm{mag}}\left( R \right) =
  \frac{\pi R^2 B_{\rm{c}}}{\left(R/R_{\rm{c}}\right)^2+1}.
  \label{app:eq:mag_flux}
\end{equation}
Thus, the dimensionless mass-to-flux ratio $\magmu$ can be written as
\begin{align}
  \magmu
  &= \left(\frac{M_{\rm{c}}}{\Phi_{\rm{mag}}\left(R_{\rm{c}}\right)}\right) \left/
  \left(\frac{M_{\rm{c}}}{\Phi_{\rm{mag}}}\right)_{\rm{crit}} \right. \nonumber \\
  &= \frac{2}{\pi}
  \left(\frac{M_{\rm{c}}}{\Phi_{\rm{mag}}}\right)_{\rm{crit}}^{-1}
  \frac{M_{\rm{c}}}{R_{\rm{c}}^2 B_{\rm{c}}}.
  \label{app:eq:mass-to-flux}
\end{align}

With the above equations (\ref{app:eq:Egrav}), (\ref{app:eq:Emag}),
and (\ref{app:eq:mass-to-flux}), the scaling relation between $\gmag$
and $\magmu$ is obtained as follow:
\begin{align}
  \gmag &= \frac{E_{\rm{mag}}}{|E_{\rm{grav}}|} \nonumber \\
  &=
  \frac{\Lambda_{\rm{mag}}}{G\Lambda_{\rm{grav}}}
  \frac{R_{\rm{c}}^4 B_{\rm{c}}^2}{M_{\rm{c}}^2} \nonumber \\
  &=
  \frac{4}{\pi^2 G}
  \left(\frac{M_{\rm{c}}}{\Phi_{\rm{mag}}}\right)_{\rm{crit}}^{-2}
  \frac{\Lambda_{\rm{mag}}}{\Lambda_{\rm{grav}}}
  \frac{1}{\magmu^2} \nonumber \\
  &\propto \magmu^{-2}.
  \label{app:eq:gmag}
\end{align}
With equation (\ref{app:eq:gmag}), ${\cal{M}_{\rm{A}}}$ is
expressed by equation (\ref{eq:alf_mach}) with the initial
parameters of ${\cal{M}_{\rm{s}}}$ and $\magmu$.

\section{
  Numerical convergence tests for the random seed of initial
  turbulence and the numerical resolution
}
\label{app:sec:numerical_convergence_test}

\begin{figure*}
  \begin{center}
    \includegraphics[clip,width=0.85\linewidth]{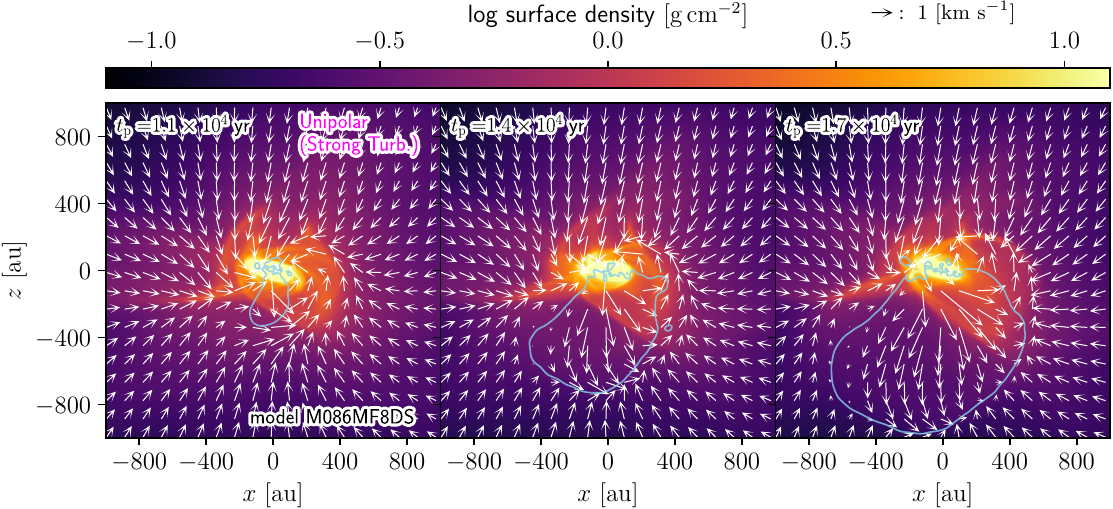}
  \end{center}
  \caption{
    Evolution of the surface density for the model M086MF8DS, which is
    identical to the model M086MF8 ($\machs=0.86$, $\magmu=8$) except
    for the random seed used to generate the initial turbulent
    velocity field.  \\
    {Alt text: Surface density evolution for the different seed model.}
  }
  \label{fig:app:surface_density_diff_seed}
\end{figure*}

As described in section \ref{subsec:init_cond}, all the models in the
main body of this paper adopt the same realization of the initial
turbulent velocity field and the same numerical resolution of
$N_{\rm{SPH}}=10^6$, following \citet{2024ApJ...963...20T}.  This
setup enables controlled comparisons among models with different
turbulent and magnetic energies, such that the differences among the
models arise solely from the adopted physical parameters.  In this
appendix, we examine the robustness of the results for the random seed
used to generate the initial turbulent velocity field and the adopted
numerical resolution.

First, we examine the dependence of the results on the random seed for
initial turbulence.  We perform an additional simulation for the model
M086MF8 ($\machs=0.86$, $\magmu=8$) using a different random seed for
the initial turbulent velocity field while keeping the physical
parameters unchanged.  Figure \ref{fig:app:surface_density_diff_seed}
shows the evolution of the surface density for this model.  As
expected from the stochastic nature of the turbulence, the launching
direction of the outflow differs from that shown in Figure
\ref{fig:surface_density_map}.  However, the unipolar outflow is
driven even with a different random seed for the initial turbulent
velocity field, indicating that the qualitative result does not depend
on a particular realization of the turbulent velocity field.

\begin{figure}
  \begin{center}
    \includegraphics[clip,width=0.95\linewidth]{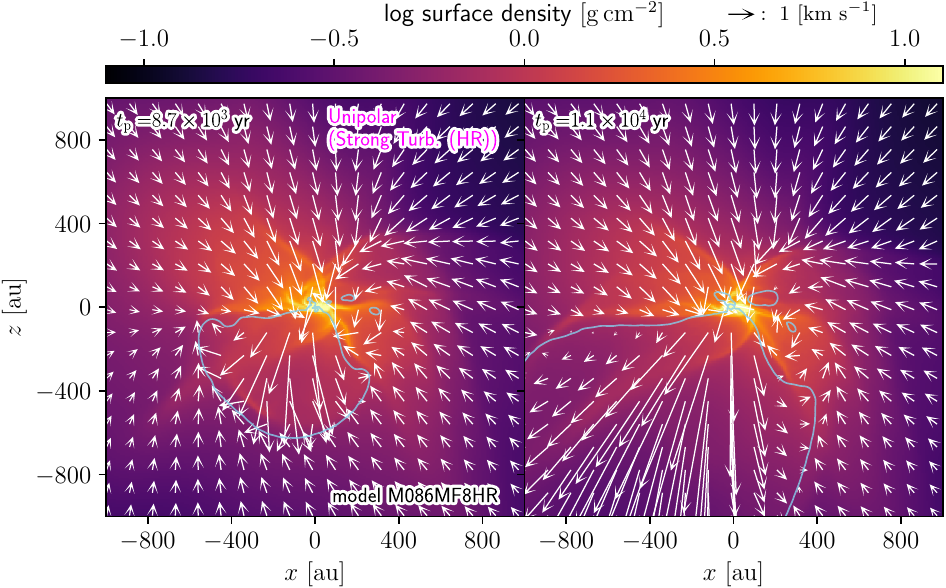}
  \end{center}
  \caption{ Evolution of the surface density for the model M086MF8HR,
    which is identical to the model M086MF8 ($\machs=0.86$,
    $\magmu=8$) except for the number of SPH particles.  The number of
    SPH particles is twice that of the model M086MF8.  \\
    {Alt text: Surface density evolution for the high-resolution model.}
  }
  \label{fig:app:surface_density_high_res}
\end{figure}

Next, we examine the dependence of the results on the numerical
resolution.  We perform additional simulations with different numbers
of SPH particles, corresponding to twice and half the fiducial
resolution of $N_{\rm{SPH}}=10^{6}$, while keeping all physical
parameters unchanged.  Figure \ref{fig:app:surface_density_high_res}
shows the evolution of the surface density for the high-resolution
model M086MF8HR, in which the number of SPH particles is increased to
$2N_{\rm{SPH}}=2 \times 10^{6}$ ($4.2\times10^{6}$ particles in the
entire computational domain of $r<10R_{\rm{c}}$).  In this
high-resolution model, the formation of the unipolar outflow is
qualitatively unchanged.

\begin{figure}
  \begin{center}
    \includegraphics[clip,width=0.95\linewidth]{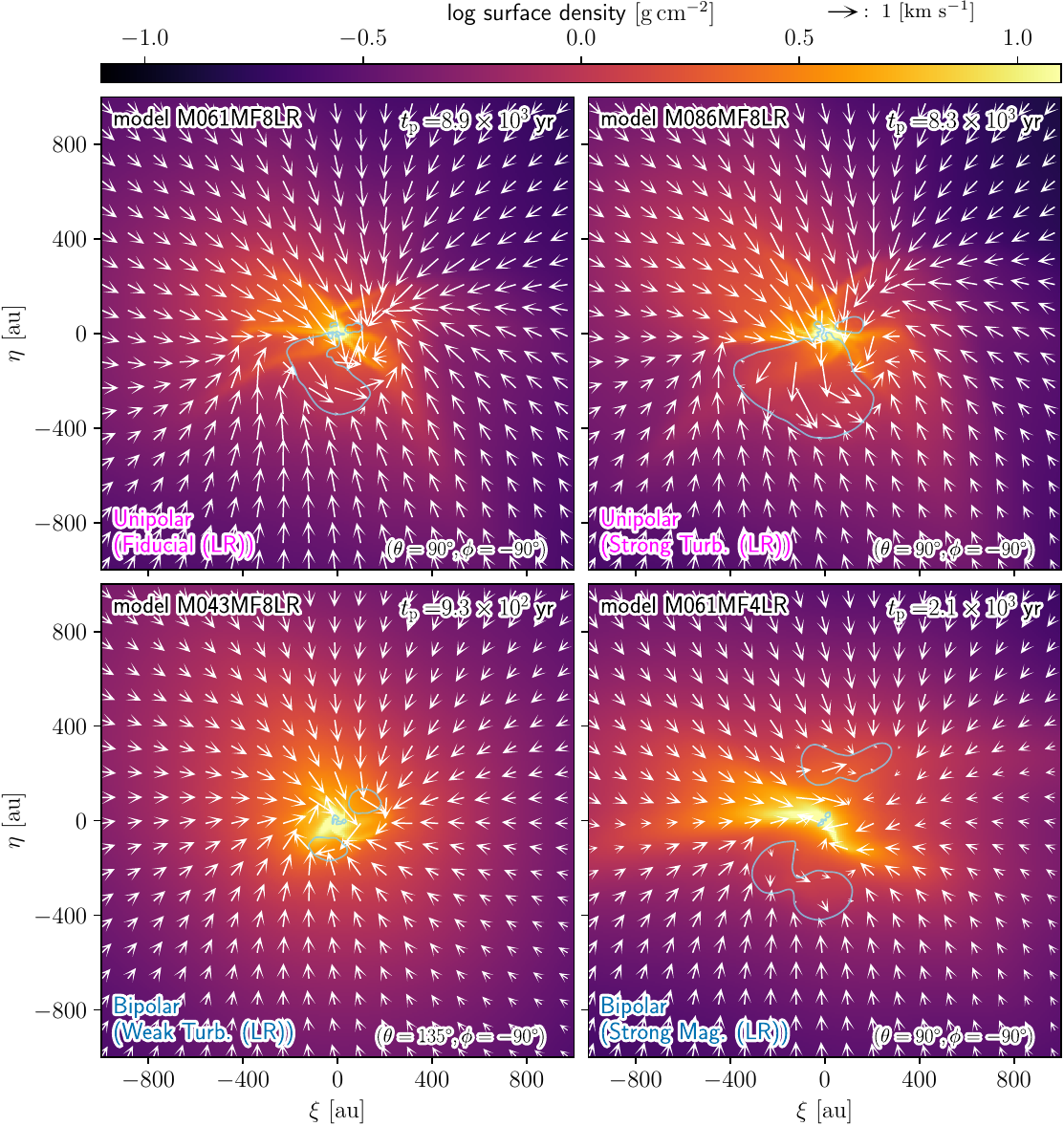}
  \end{center}
  \caption{ Evolution of the surface density for the low-resolution
    models.  The models correspond to those shown in Figure
    \ref{fig:surface_density_map}, but with the number of SPH
    particles reduced to $N_{\rm{SPH}}/2$.  \\
    {Alt text: Surface density evolution for the low resolution models.}
  }
  \label{fig:app:surface_density_low_res}
\end{figure}

Figure \ref{fig:app:surface_density_low_res} shows the evolution of
the surface density for the low-resolution models.  The physical
parameters are identical to those of the four models shown in Figure
\ref{fig:surface_density_map}, except that the number of SPH particles
is reduced to $N_{\rm{SPH}}/2=5\times 10^5$.  The outflow morphologies
are qualitatively identical to those of the fiducial models.

Therefore, the qualitative outcome remains unchanged for different
random seed for initial turbulence and for both the high-resolution
and low-resolution simulations, indicating that the main conclusions
of this paper are robust against the random seed and numerical
resolution explored in this study.

\section{Spherical harmonics}
\label{app:spherical_harmonics}

\begin{figure*}
  \begin{center}
    \includegraphics[clip,width=0.95\linewidth]{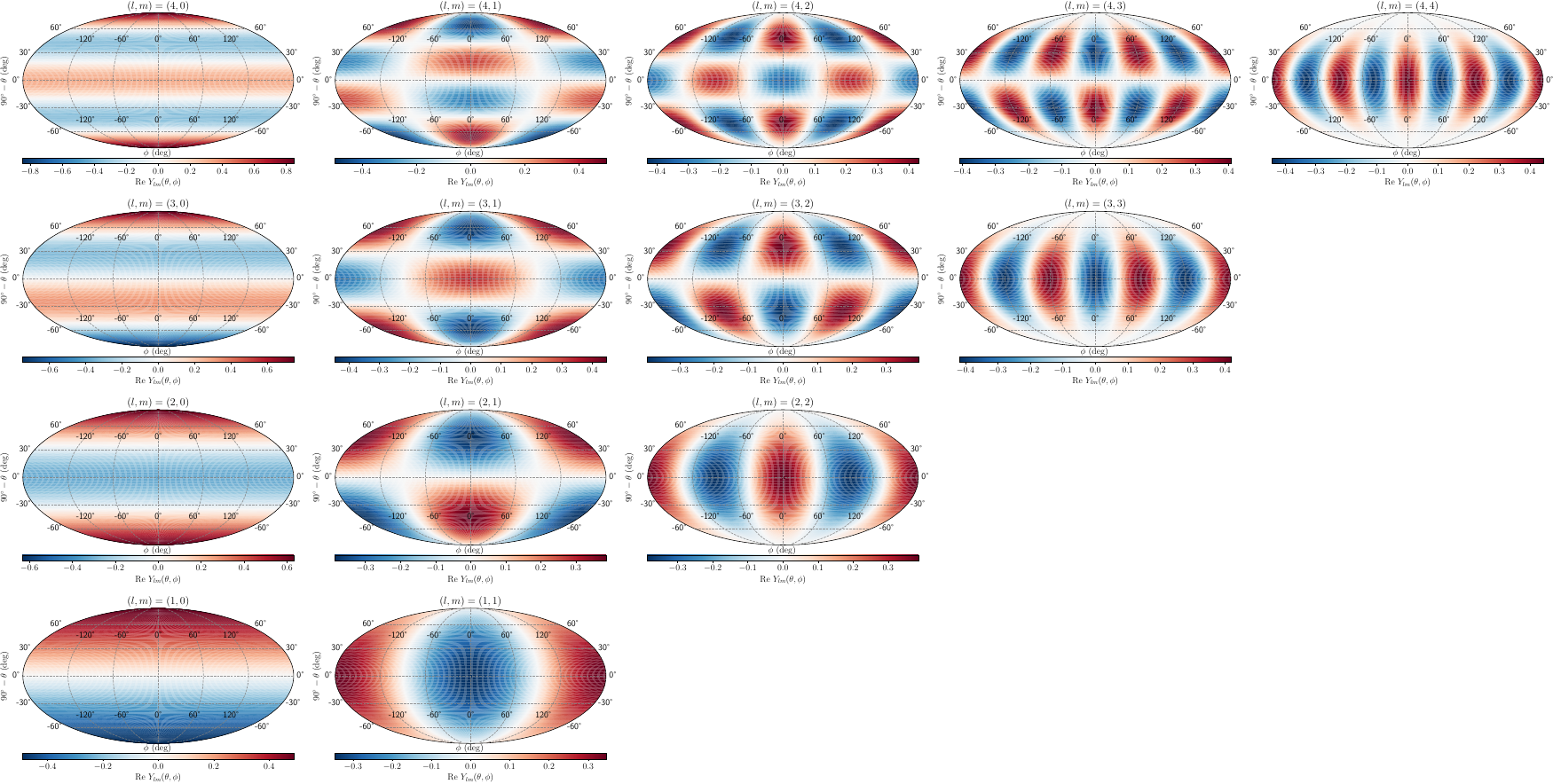}
  \end{center}
  \caption{
    Examples of the real part of $Y_{l}^{m}(\theta,\phi)$
    for $l=1,\thinspace 2,\thinspace 3,\thinspace 4$ and the corresponding values of
    $m=0$ to $l$.  \\
    {Alt text: Example maps of the spherical harmonics.}
  }
  \label{fig:app:Y_lm_real}
\end{figure*}

In section \ref{subsec:polarity}, we apply the spherical
harmonics expansion for the characterization of the polarity of
outflows, which is often used in the analysis of mapping the
anisotropies in the CMB radiation (e.g., \cite{1992ApJ...396L...1S,
  2003ApJS..148....1B, 2008cosm.book.....W, 2011ApJS..192...18K,
  2014A&A...571A...1P, 2020A&A...641A...1P, 2020A&A...641A...6P}).
This appendix briefly reviews the analysis of the spherical harmonics
expansion used in this study.

The spherical harmonics $Y_{l}^{m}(\theta,\phi)$ are a set of
orthogonal functions defined on the surface of a sphere as
\begin{align}
  Y_{l}^{m}(\theta,\phi)
  &\equiv
  (-1)^m
  \sqrt{ \frac{(2l+1)}{4\pi} \frac{(l-m)!}{(l+m)!} }
  P_{l}^{m}(\cos\theta) \exp(i m \phi),
  \label{app:eq:Y_lm}
\end{align}
where $P_{l}^{m}(x)$ are the associated Legendre functions defined
with the Legendre polynomials $P_l(x)$ as
\begin{align}
  P_l^m(x) &\equiv  (1 - x^2)^{m/2} \frac{d^m}{dx^m} P_l(x)
  ~~{\rm{for}}~m \geq 0,
  \label{app:eq:P_lm} \\
  P_l(x) &\equiv \frac{1}{2^l l!}\frac{d^l}{dx^l}\left\{(x^2-1)^l\right\}.
  \label{app:eq:P_l}
\end{align}
$Y_{l}^{m}(\theta,\phi)$ are characterized by the two parameters $l$
and $m$, which take values $l=0,1,2,\ldots$ and
$m=-l,-l+1,\ldots,l-1,l$.  The definitions here follow those written
in \citet{1962qume.book.....M}.  Since the associated Legendre
functions satisfy the relation of
\begin{align}
  P_l^{-m}(x) = (-1)^{m} \frac{(l-m)!}{(l+m)!} P_l^{m}(x),
  \label{app:eq:P_l-m}
\end{align}
the following holds:
\begin{align}
  Y_{l}^{m}(\theta,\phi)^{*} = (-1)^{m} Y_{l}^{-m}(\theta,\phi).
  \label{app:eq:Y_lm*}
\end{align}
The spherical harmonics are orthonormal as
\begin{align}
  \int_{-1}^{1} d(\cos\theta)
  \int_{0}^{2\pi} d\phi
  Y_{l}^{m}(\theta,\phi) Y_{l^{'}}^{m^{'}}(\theta,\phi)^{*}
  = \delta_{l{l^{'}}} \delta_{m{m^{'}}},
  \label{app:eq:orthonormal_Y_lm}
\end{align}
where $\delta_{ij}$ is the Kronecker delta.  We present some examples
for $Y_{l}^{m}(\theta,\phi)$ in Figure \ref{fig:app:Y_lm_real}.

The spherical harmonics expansion for the function
$f(\nhat)=f(\theta,\phi)$ are described as follows:
\begin{align}
  f(\nhat) = {\ssum_{l=0}^{\infty}} {\ssum_{m=-l}^{m=l}}
  a_{lm} Y_{l}^{m}(\theta,\phi),
  \label{app:eq:sph_harm_expansion}
\end{align}
where $a_{lm}$ are the expansion coefficients calculated as
\begin{align}
  a_{lm} = \int_{4\pi} f(\nhat)
  Y_{l}^{m}(\theta,\phi)^{*} d\Omega.
  \label{app:eq:a_lm}
\end{align}
In this paper, we set $f(\nhat)=\Delta \hat{P}_{\rm{out}}(\nhat)$.  By
the definition of equation (\ref{eq:delta_mom_out}), the following
integral is satisfied:
\begin{align}
  \int_{4\pi} \Delta \hat{P}_{\rm{out}}(\nhat) d\Omega = 0.
  \label{app:eq:integral_deltaP}
\end{align}
Thus, the monopole term $a_{00}$ is calculated as
\begin{align}
  a_{00} &= \int_{4\pi} \Delta \hat{P}_{\rm{out}}(\nhat) Y_{0}^{0}(\theta,\phi)^{*} d\Omega
  \nonumber \\
  &= \frac{1}{\sqrt{4\pi}} \int_{4\pi} \Delta \hat{P}_{\rm{out}}(\nhat) d\Omega
  \nonumber \\
  &= 0.
  \label{app:eq:a00}
\end{align}
The expansion coefficients $a_{lm}$ satisfy the following relation:
\begin{align}
  a_{lm}^{*} = (-1)^{m} a_{l-m}.
  \label{app:eq:a_lm*}
\end{align}
Using equations (\ref{app:eq:a_lm*}), the following equations are
valid:
\begin{align}
  a_{lm} a_{lm}^{*} = a_{l-m} a_{l-m}^{*}.
  \label{app:eq:|a_lm|}
\end{align}
Therefore, we plot $|a_{lm}|$ for $m\geq0$ in Figure
\ref{fig:polarity_mom_alm}.

\bibliographystyle{aasjournal}
\bibliography{export-bibtex, export-bibtex-2, export-bibtex-3,
  takaishids_article-bibtex} 

\end{document}